\documentclass[11pt]{article}

\usepackage[breakable]{tcolorbox}
\usepackage{amssymb}
\usepackage{rotating}
\usepackage[greek,english]{babel}
\usepackage{graphicx}
\usepackage{caption}
\usepackage{lscape}
\usepackage{rotating}
\usepackage{epstopdf}
\usepackage{algorithm}
\usepackage{color}

\input{epsf}

\newcounter{saveeqn}

\newcommand{\bfT}{\mbox{\boldmath $T$}}

\begin{document}

\parindent0mm \parskip0.6cm


{\LARGE {\bf Exploring complex dependence structures using Bayesian bi-clustering and log-linear graphical modelling} }

\begin{large}
Michail Papathomas$^{*}$
\end{large}
\\ {\it $^{*}$ Corresponding author: School of Mathematics and Statistics,  University of St Andrews, UK, The Observatory, Buchanan Gardens, St Andrews, KY16 9LZ, \\ M.Papathomas@st-andrews.ac.uk, 0044 1334461818}

\vspace{0.3cm}

{\bf ABSTRACT: Bayesian partitioning is utilised simultaneously on subjects and categorical variables to reveal complex dependence structures. Clusters of variables are referred to as views. Variable selection highlights the variables that drive the clustering of the subjects within each view. We derive theoretical results on the relation between the variables' dependence structure and the inferences derived from bi-clustering. The results relate to marginal independence and conditional independence. Using simulated and real data, we demonstrate the applicability of bi-clustering results in assisting log-linear graphical model determination, leading to the efficient exploration of typically vast model spaces. This work sheds light on the relation between two very different but equally popular Bayesian approaches; mixture modelling, that benefits from a large number of variables, and graphical log-linear modelling, which describes explicitly the variables' dependence structure and allows to evaluate the uncertainty associated with model determination. 
}

{\it Key words:} Bayesian model selection, sparse contingency tables, graphical models

\section{Introduction}

Inferring the dependence structure between a number of variables is crucial for understanding the manner in which they interrelate, impact each other and form possible pathways. For example, in multimorbidity studies, chronic medical conditions co-occur within populations (Jensen et al. 2014). In epidemiology, it is of interest to investigate risk factors for antibiotic resistance holistically, by combining microbiology, patient, household and community level data (Keenan et al. 2024).

This manuscript concerns categorical variables. We demonstrate that {\it simultaneous} Bayesian partitioning on subjects and variables (referred to as bi-clustering), combined with variable selection, can be part of an efficient algorithm that reveals complex dependence structures.  Variables are partitioned into views while, within each view, subjects are partitioned into clusters. We adopt flexible Bayesian clustering because it allows for evaluation of the uncertainty through the rich MCMC output. We implement finite mixture modelling (Gr{\"u}n and Malsiner-Walli, 2022), allowing for empty clusters and views, as it provides with valid and plausible inferences with standard sampling approaches. Other options include mixtures of finite mixtures (Fr{\"u}hwirth-Schnatter, Walli and Gr{\"u}n, 2021) or Dirichlet process mixture models (Dunson and Xing, 2009). 
Variable selection is integrated within the model, implementing the modified variable selection step described in Papathomas et al. (2012). This identifies variables that contribute substantially to the clustering of the subjects within each view. Note that the bi-clustering model is different to multi-view clustering (Franzolini et al.,2026; Dombowsky and Dunson, 2026) as the variables are not grouped into known views or domains. Variable allocations into views are estimated, similarly to the allocation of subjects into clusters within each view. 

We utilise Bayesian graphical log-linear model determination to evaluate the uncertainty associated with the variables' dependence structure, assisted by the output from bi-clustering. Log-linear modelling is the default approach for estimating the dependence structure among categorical variables. The search of a vast model space with algorithms such as the Reversible Jump (Green, 1995) is computationally challenging. For instance,  20 categorical variables with 2 levels each imply a contingency table with $2^{20}$ cells and approximately $1.5\times 10^{57}$ possible log-linear models. Increasing the efficiency of the process by quickly locating areas of high probability is crucial. 

Models that combine clustering and variable selection highlight variables because they combine together to create homogenous clusters. It is therefore expected that this modelling could potentially determine variables that relate to each other and the manner in which this happens. In this manuscript, we show that inferences from bi-clustering inform on the existence of edges present in a log-linear graphical model. This is a major departure from the single-view approach of Papathomas and Richardson (2016), where the derived theoretical results led to consistent detection of variables that are independent of all others. Here, theoretical results concern independence between groups of variables and also conditional independence. They allow to derive pseudo-probabilities for the existence of log-linear graphical model edges by post processing the MCMC output from the bi-clustering partition model. The probabilities are included in a matrix that informs on the possible structure of a likely graphical log-linear model. This matrix is fed to formal log-linear graphical model search that determines posterior probabilities for the models that comprise the typically large model space. The proposed model search identifies high probability models in fewer iterations compared to a less informed approach, increasing the efficiency of the algorithm.  

For other studies relevant to the relation between mixture and log-linear modelling see Dunson and Xing (2009), Bhattacharya and Dunson (2012), Marbac et al. (2014), Johndrow et al. (2014) and Zhou et al. (2015). Mixture models are used extensively for the clustering of network populations; see, for example, Mantziou et al. (2024) or D'Angelo et al. (2023). In contrast to the modelling proposed here, these manuscripts concern data from multiple networks, without implementing variable selection. In addition, no reversible jump model search algorithm is utilised for the detection of high probability graphical models. 

In Section 2, we provide a detailed description of the bi-clustering model and integrated variable selection approach, while Section 3 provides a brief account of the more familiar log-linear modelling. Section 4, contains theoretical results on the correspondence between marginal independence and conditional independence on one hand, and variable selection within bi-clustering on the other. The proposed model search approach for log-linear models is described within Section 4 too. Simulated data sets are analysed in Section 5, and two real data sets in Section 6. A discussion concludes the manuscript.

\section{The bi-clustering Bayesian mixture model}

For subject $i$, $i=1,...,n$, a  profile $x_{i}$ is a vector of categorical variable values $x_{i}=(x_{i1},...,x_{iP})$, where $P$ is the number of variables. We denote by $x_{.p}$ the $p^{th}$ categorical random variable, $p=1,...,P$. Variables are clustered into views, so that the clustering of the subjects can be different within each view; see Kirk et al. (2023). Let $s_p=v$ denote that $x_{.p}$ belongs to view $v$, $v=1,...,V$. Let also,  
\[
P(s_{p}=v |\xi) = \xi_{v},
\] 
where $\xi=(\xi_{1},...,\xi_{V})$. Let $z^{v}=\{z^{v}_{1},...,z^{v}_{n} \}$, where $z^{v}_{i}$ is an allocation variable, so that $z^{v}_{i}=c^{v}$
denotes that subject, $i$, belongs to cluster $c^{v}$ considering view $v$. Here, $c^{v}=1,...,C^{v}$, where $C^{v}$ is the maximum number of clusters for view $v$. Let, 
\[
P(z_{i}^{v}=c^{v} | \psi) = \psi_{c^{v}}^{v},
\]
where, $\psi$ is an irregular matrix that contains the elements $\psi_{c^{v}}^{v}$. For $C^{1}=...=C^{V}=C$, $\psi$ is a $V \times C$ matrix. 
Denote by $\phi^{c^{v}}_{p}(x)$ the probability that the $p^{th}$ variable $x_{.p}$, allocated to view $v$, is equal to $x$, when the subject belongs to cluster $c^{v}$.
Given that $z^{v}_{i}=c^{v}$, variable $x_{.p}$ has a multinomial distribution with
cluster specific parameters
$\phi^{c^{v}}_{p}=[\phi^{c^{v}}_{p}(1),...,\phi^{c^{v}}_{p}(M_{p})]$. Here,
$M_{p}$ denotes the number of categories of $x_{.p}$. 
We assume that, a priori, $\phi^{c^{v}}_{p}\sim \mbox{Dirichlet}(\lambda_{1},...,\lambda_{M_{p}})$. Let $\phi=\{\phi^{c^{v}}_{p}, \hspace{0.1cm} v=1,...,V, c^{v}=1,...,C^v, p=1,...,P \}$.
We adopt a flexible truncated stick-breaking prior on the allocation weights
$\psi$, with a random parameter $\alpha$ (West, 1992; Ishwaran and James, 2001), as shown in the description of the model below. 
\[
x_i | z,\phi \sim \prod_{p=1}^{P} \phi^{z^{v}_{i}}_{p}(x_{ip}) \mbox{ for } i=1,2,...,n, \mbox{ and } v=1,2,...,V
\]
\[
\phi^{c^{v}}_{p}(x_{ip}) \sim \mbox{Dirichlet}(\lambda_{1},...,\lambda_{M_{p}}) \mbox{ for }  c=1,2,...,C
\]
\[
P(z^{v}_{i}=c^{v}|\psi)=\psi^{v}_{c^{v}} 
\]
\[
\psi^{v}_{c^{v}}=w^{v}_{c^v} \prod_{l<c^{v}} (1-w^{v}_{l}) \mbox{ for } c^{v}=2,...,C^{v}, \mbox{ with } \psi^{v}_{1} = w^{v}_{1}, \mbox{ and } w^{v}_{C^{v}}=1 
\]
\[
w^{v}_{c^v} | \alpha \sim \mbox{Beta}(1,\alpha) \mbox{ for } c^{v}=1,...,C^{v}-1
\]
\[
\alpha \sim U(0,10)
\]
\[
\xi_{v}=u_{v} \prod_{l<v} (1-u_{l}) \mbox{ for } v=2,3,...,V 
\]
\[
\xi_{1} = u_{1}, \mbox{ and } u_{V}=1,
\]
\[
u_{v} | \alpha_{\xi} \sim \mbox{Beta}(1,\alpha_{\xi}) \mbox{ for } v=1,...,V-1
\]
\[
\alpha_{\xi} \sim U(0,10)
\]

Alternatively, for $\psi^{v} = \{\psi^{v}_{1},...,\psi^{v}_{C^v} \}$ one can assume Dirichlet priors for the $\psi$ and $\xi$ parameters so that, 
\[
\psi^{v} | \alpha \sim \mbox{Dirichlet}(\alpha,...,\alpha), 
\]
i.e.,
\[
P(\psi^{v}|\alpha) = \prod_{c^{v}=1}^{C^{v}} \left( \psi_{c^{v}}^{v} \right)^{\alpha-1},
\]
and
\[
\xi | \alpha_{\xi} \sim \mbox{Dirichlet}(\alpha_{\xi},...,\alpha_{\xi}),
\]
i.e.,
\[
P(\xi|\alpha_{\xi}) = \prod_{v=1}^{V} \left( \xi_{v} \right)^{\alpha_{\xi}-1}.
\]

This implies the following mixture model for the likelihood of the observations,

\[
P(x_{i} | \phi,\psi,\xi) = 
\sum_{v_{1}=1}^{V} ... \sum_{v_{P}=1}^{V} \sum_{c^{1}=1}^{C^{1}} ...
\sum_{c^{V}=1}^{C^{V}} 
\]
\[
P(x_{.1}=x_{i1},...,x_{.P}=x_{iP} | 
z_{i}^{1}=c^{1},...,z_{i}^{V}=c^{V},s_{1}=v_{1},...,s_{P}=v_{P},\phi,\psi,\xi)
\]
\[
\times 
P(z_{i}^{1}=c^{1},...,z_{i}^{V}=c^{V} | s_{1}=v_{1},...,s_{P}=v_{P},\phi,\psi,\xi) 
\]
\[
= \sum_{v_{1}=1}^{V} ... \sum_{v_{P}=1}^{V} \sum_{c^{1}=1}^{C^{1}} ...
\sum_{c^{V}=1}^{C^{V}} 
\]
\[
\prod_{p:v_{p}=1} \phi_{p}^{c^{1}}(x_{ip}) \times ... \times 
\prod_{p:v_{p}=V} \phi_{p}^{c^{V}}(x_{ip})
\]
\[
\times \psi_{c^{1}}^{1} \times ... \times  \psi_{c^{V}}^{V} \times
\xi_{v_{1}} \times ... \times \xi_{v_{P}}.
\]

Note that, for given $s_{1}=v_{1},...,s_{P}=v_{P}$, 

\begin{eqnarray}
P(x_{.1}=x_{i1},...,x_{.P}=x_{iP} | s_{1} = v_{1},...,s_{P}=v_{P}, \phi, \psi, \xi) \nonumber \\
= \sum_{c^{1}=1}^{C^{1}} ...
\sum_{c^{V}=1}^{C^{V}} 
\prod_{p:v_{p}=1} \phi_{p}^{c^{1}}(x_{ip}) \times ... \times 
\prod_{p:v_{p}=V} \phi_{p}^{c^{V}}(x_{ip}) 
\times \psi_{c^{1}}^{1} \times ... \times  \psi_{c^{V}}^{V}.
\end{eqnarray}

The variable selection approach that identifies the variables that are important for the formation of clusters within views follows Papathomas et al. (2012).
We assume cluster specific binary indicators, $\gamma_{p}^{c^{v}}$, so that $\gamma_{p}^{c^{v}}=1$ when variable $x_{.p}$ is important for allocating subjects to cluster $c^{v}$, considering the view $v$ where $x_{.p}$ is allocated; otherwise $\gamma_{p}^{c^{v}}=0$.
Denote by $\pi_{p}(x_{ip})$ the marginal probability that variable $x_{.p}$ takes the value $x_{ip}$,  $P(x_{.p}=x_{ip})$. 
The probability that variable $x_{.p}$ is observed as $x_{ip}$, when subject, $i$, belongs to cluster $c$, is written as,
\begin{eqnarray}
P(x_{.p}=x_{ip} \mid z^{v}_{i}=c^{v})=[\phi^{c^{v}}_{p}(x_{ip})]^{\gamma^{c^{v}}_{p}} \times [\pi_{p}(x_{ip})]^{(1-\gamma^{c^{v}}_{p})}.
\end{eqnarray}
 Now, we can write,  
\[
\pi_p(x_{ip})=P(x_{.p}=x_{ip}) = \sum_{c^{v}} \psi_{c^{v}}  [\phi^{c^{v}}_{p}(x_{ip})]^{\gamma^{c^{v}}_{p}} \times [\pi_{p}(x_{ip})]^{(1-\gamma^{c^{v}}_{p})}.
\]
We assume that the $\gamma_{p}^{c^{v}}$ are independent Bernoulli
variables with $\gamma_{p}^{c^{v}} \sim \mbox{Bernoulli} (\rho_{p})$, $0<\rho_{p}<1$. Here, $\rho_p$ describes the probability that variable $x_{.p}$ is important for the partitioning of the subjects in relation to the entire process, rather than for a specific cluster only. 
For $\rho_p$, one may choose to consider a sparsity inducing prior with an atom at zero, so that
$\rho_{p} \sim 1_{\{f_{p}=0\}} \delta_{0}(\rho_{p}) + 1_{\{f_{p}=1\}}  \mbox{Beta} (\alpha_{\rho}, \beta_{\rho})$,
where $f_{p} \sim \mbox{Bernoulli}(0.5)$. We do not consider this option in this manuscript, as the number of variables we examine does not exceed 10. 
The model described in this Section is fitted using custom R code, available in the form of two R packages, one for bi-clustering and another for log-linear model search. The full conditional distributions for the MCMC sampler are given in the Supplemental material, Section S1.

\section{Log-linear graphical models}

Log-linear models represent the dependence structure between categorical variables. Importantly, they allow to evaluate the uncertainty associated with this structure by formal Bayesian model comparison. Let $\mathcal{P}$ denote the finite set of the $P$ categorical variables. The resulting data can be arranged as counts in a $P$-way contingency table. A Poisson log-linear interaction model is a generalized linear model where the data are the cell counts of the contingency table; see the Supplemental material, Section S2, for a formal definition of an interaction term in a log-linear model. The number of all possible log-linear models is $2^{(2^{P})}$. Graphical models are represented by a graph where each node (or vertex) is an element of $\mathcal{P}$. Two nodes may be connected by an edge. Nodes not connected directly by an edge are independent conditionally on all other nodes (pairwise Markov property). Conditionally on nodes to which $x_{.p}$ is directly connected, $x_{.p}$ is independent of all other nodes (local Markov property). Two sets of nodes are independent when they are separated by another set, conditionally on the separating set (global Markov property); see Lauritzen (2011) for a more rigorous definition of Markov properties. The number of possible graphical models is $2^{H}$, where $H=P!/(2(P-2)!)$,
assuming the intercept and all factor main effects are included in the model.

\section{Bi-clustering outcomes and dependence}

\subsection{Independence and conditional independence}

Theorems 1 and 2 allow to detect independent groups of random variables. The remaining Theorems relate to conditional independence. 

{\bf Theorem 1:} If random variables $\{x_{.p_{1}},\dots,x_{.p_{2}} \}$ are allocated to view $v$, while variables $\{x_{.q_{1}},\dots,x_{.q_{2}} \}$ are allocated to a different view $v^{'}$, then $\{x_{.p_{1}},\dots,x_{.p_{2}} \}$ are independent of $\{x_{.q_{1}},\dots,x_{.q_{2}} \}$. 

{\it Proof:} See Appendix.

{\bf Theorem 2:} Denote by $v_{p}$ the view variable $x_{.p}$ is allocated to. A set of random variables $\{x_{.p_{1}},\dots,x_{.p_{2}} \}$ is independent of another set $\{x_{.q_{1}},\dots,x_{.q_{2}} \}$ if and only if $v_{p}\neq v_{q}$ for all $p \in \{p_{1},...,p_{2} \}$ and $q \in \{q_{1},...,q_{2} \}$.

{\it Proof:} See Appendix.

Assume that within some view $v$, the subjects are grouped into $C^{v}$ clusters. Let $\Gamma$ be a subset of $\{1,\dots, C^{v} \}$. Denote by 
$\Gamma^{\complement}$ the compliment of $\Gamma$ with respect to $\{1,\dots, C^{v} \}$. Without any loss of generality, consider that, for two variables $x_{.p}$ and $x_{.q}$ in that view, \\
 $\gamma_{p}^{c}=0$, $\gamma_{q}^{c}=1$, for $c \in \Gamma_1$, \\
 $\gamma_{p}^{c}=1$, $\gamma_{q}^{c}=0$, for $c \in \Gamma_2$, \\
 $\gamma_{p}^{c}=1$, $\gamma_{q}^{c}=1$, for $c \in \Gamma_3$, \\
 $\gamma_{p}^{c}=0$, $\gamma_{q}^{c}=0$, for $c \in \Gamma_4
=(\Gamma_1 \cup \Gamma_2 \cup \Gamma_3)^{\complement}$. 

{\bf Proposition 1:} For two variables $x_{.p}$ and $x_{.q}$ in the same view, it is not possible that the view only comprises clusters of type $\Gamma_1$ or $\Gamma_2$ or both, without any cluster in the $\Gamma_3$ or  $\Gamma_4$ category. 

{\it Proof:} See Appendix.

{\bf Theorem 3:} For two random variables $x_{.p}$ and $x_{.q}$ within the same view $v$, if only clusters of type $\Gamma_3$ are present, then the two variables are not conditionally independent. 

{\it Proof:} See Appendix.
 
{\bf Theorem 4:} For two random variables $x_{.p}$ and $x_{.q}$ within the same view $v$, if the view contains only clusters of type $\Gamma_1$ and $\Gamma_3$, or only clusters of type $\Gamma_2$ and $\Gamma_3$, then the two variables are conditionally independent. 

{\it Proof:} See Appendix.

The following notation is relevant to Theorem 5 below, and will also be used later within the manuscript. Consider two random variables $x_{.p}$ and $x_{.q}$ within the same view $v$. 
To simplify the notation, and without loss of generality, assume that the rest of the variables in this view are $(x_{.1}, ... , x_{.r})$. Let,  
\[
f_{rest}(\phi,\pi,c) = \left( (\phi_1^c)^{\gamma_1^c} \pi_{1}^{1-\gamma_1^c} \right) \times ... \times 
\left( (\phi_r^c)^{\gamma_r^c} \pi_{r}^{1-\gamma_r^c} \right).
\]

{\bf Theorem 5:} Consider two random variables $x_{.p}$ and $x_{.q}$ within the same view $v$. Assume that the rest of the variables in this view are $(x_{.1}, ... , x_{.r})$. 
Assuming that only clusters of types $\Gamma_1$, $\Gamma_2$ and $\Gamma_3$ are present, if 
\[
\left( \sum_{c_1 \in \Gamma_1} \psi_{c_1} f_{rest}(\phi,\pi,c_1) (\phi_{q}^{c_1} - \pi_q) \right) 
\times \left( \sum_{c_2 \in \Gamma_2} \psi_{c_2} f_{rest}(\phi,\pi,c_2) (\pi_p - \phi_{p}^{c_2}) \right) = 0, 
\]
then the two variables are conditionally independent. 

{\it Proof:} See Appendix.

\subsection{Summary of the bi-clustering outcomes}

The results in Section 4.1 provide a blueprint for building a summary matrix $\bfT$ that translates the bi-clustering output into information that is relevant to graphical log-linear model selection. Matrix $\bfT$ is constructed in such a manner so that if element $t_(p_{1},p_{2})$, $1\leq p_1 < p_2\leq P$, is zero or close to zero, this implies that an edge between $x_{.p_1}$ and $x_{.p_2}$ is not likely to be present in a highly supported log-linear graphical model.  

At this stage, we should point out that graphical log-linear modelling does not describe all possible dependence structures between categorical variables. For example, the set of all possible hierarchical models is larger than the set of possible graphical models. In addition, the Theorems in Section 4.1 do not cover all eventualities in terms of the types of clusters that can be present within a view. Finally, there is random variation within the MCMC sampler, and mixing/convergence considerations; see Chaumeny et al. (2022).  This is why the derived theoretical results do not provide an exact template for a direct translation of the bi-clustering output into inferences on the variables' dependence structure. The proposed construction below takes this into account by allowing for some variation within the bi-clustering output and setting thresholds for deciding on the presence or absence of different types of cluster. 

For two variables $x_{.p}$ and $x_{.q}$, denote by $|\Gamma_l|$ the cardinality of the set of clusters of type $\Gamma_l$, $l=1,2,3,4$, within view $v$. Let $thres$ take a value within $\{0,1,...,C\}$. We assume that a type of cluster is not present within a view if $|\Gamma_l|\leq thres$. So, in practice, $thres$ should take some small non-zero value, and we suggest $thres=1$ to allow for a small departure from $|\Gamma_l|= 0$ when clusters of type $l$ are not present. Let also $sumthres$ be a small positive number. This is a threshold that will allow for a small departure from zero for the sums described within Theorem 5. We further discuss the choice of $sumthres$ in Section 5.2. 

Consider two variables $x_{.p}$ and $x_{.q}$ in view $v$. As before, assume that the rest of the variables in this view are $(x_{.1}, ... , x_{.r})$. Ignoring the view specific indicators to simplify the notation, let,
\[
\Sigma_{\Gamma_l} = \sum_{c \in \Gamma_l} \psi_{c} f_{rest}(\phi,\pi,c) (\phi_q^c - \pi_q).
\]
The rules for the formation of the summary matrix $\bfT$ are given below. In brackets, we show the theoretical result that motivated each rule. Empirical studies informed on three additional rules that limit the number of false negative indications on the presence of an edge.  

\begin{tcolorbox}

\begin{itemize}

\item For iteration $i_t$ form matrix $\bfT^{i_t}$, so that element $t_{(p_{1},p_{2})}$ on row $p_{1}$ and column $p_{2}$, $1\leq p_{1} < p_{2}\leq P$, is initially set to zero. All other matrix cells are empty. 

\item $t_{(p_{1},p_{2})}=0$ when $x_{.p_{1}}$ and $x_{.p_{2}}$ belong to different views. (Theorems 1,2.)

\item When $x_{.p_{1}}$ and $x_{.p_{2}}$ belong to the same view, $v$:

\begin{itemize}

\item $t_{(p_{1},p_{2})}=1$ when $|\Gamma_1| \leq thres$ and $|\Gamma_2| \leq thres$ and $|\Gamma_3| \geq |\Gamma_4|$ and $|\Gamma_3| \geq thres$ (Theorem 3)

\item $t_{(p_{1},p_{2})}=1$ when $|\Gamma_1| > thres$ and $|\Gamma_2| > thres$ and $|\Gamma_3| \geq thres$ and $|\Gamma_4| \geq thres$ (Reduce false negatives)

\item $t_{(p_{1},p_{2})}=1$ when $|\Gamma_1| \leq thres$ and $|\Gamma_2| \geq thres$ and $|\Gamma_3| \geq thres$ and $|\Gamma_4| \geq thres$ (Reduce false negatives)

\item $t_{(p_{1},p_{2})}=1$ when $|\Gamma_1| \geq thres$ and $|\Gamma_2| \leq thres$ and $|\Gamma_3| \geq thres$ and $|\Gamma_4| \geq thres$ (Reduce false negatives)

\item $t_{(p_{1},p_{2})}=0$ when ($|\Gamma_1| \geq thres$ or $|\Gamma_2| \geq thres$) and $|\Gamma_3| < |\Gamma_4|$ and $|\Gamma_3| \leq thres$ and $|\Gamma_4| \leq thres$ (Proposition 1)

\item $t_{(p_{1},p_{2})}=0$ when $|\Gamma_1| \geq thres$ and $|\Gamma_2| \geq thres$ and $|\Gamma_3| \geq thres$ and $|\Gamma_3| \geq |\Gamma_4|$ and $|\Gamma_4| \leq thres$ and  
         ($|\Sigma_{\Gamma_1}| < sumthres \mbox{ or } |\Sigma_{\Gamma_2}| < sumthres$) (Theorem 5)
				
\item $t_{(p_{1},p_{2})}=1$ when $|\Gamma_1| \geq thres$ and $|\Gamma_2| \geq thres$ and $|\Gamma_3| \geq thres$ and $|\Gamma_3| \geq |\Gamma_4|$ and $|\Gamma_4| \leq thres$ and 
         ($|\Sigma_{\Gamma_1}| > sumthres \mbox{ and } |\Sigma_{\Gamma_2}| > sumthres$) (Theorem 5)

\item $t_{(p_{1},p_{2})}=0$ when $|\Gamma_1| < thres$ and $|\Gamma_2| \geq thres$ and $|\Gamma_3| \geq thres$ (Theorem 4)

\item $t_{(p_{1},p_{2})}=0$ when $|\Gamma_1| \geq thres$ and $|\Gamma_2| < thres$ and $|\Gamma_3| \geq thres$ (Theorem 4)

\item $t_{(p_{1},p_{2})}=0$ in any other case.

\end{itemize}

\item Sum up all matrices $\bfT^{i_t}$, to create an information matrix $\bfT$ as a summary of all $\bfT^{i_t}$ matrices into one.

\item For ease of interpretation, re-weight the elements of $\bfT$ by dividing all elements with the maximum element, so that the resulting matrix has elements that take values within $[0,1]$ and the maximum element is one.

\end{itemize}

\end{tcolorbox}

\subsection{The proposed log-linear model search algorithm}

The proposed model comparison approach is based on the Reversible Jump MCMC algorithm (Green, 1995) implemented in Papathomas et al. (2011). Following Papathomas and Richardson (2016), to propose the addition of an edge to the currently accepted model, we consider the elements of $\bfT$ that correspond to pairs of variables not currently connected with an edge, transform so that they sum to one, and sample an edge using the derived probabilities. To suggest an edge for removal, we consider the elements of $\bfT$ that correspond to pairs of variables already connected with an edge, transform so that they sum to one, and sample an edge using complimentary probabilities. To choose one edge to replace another, we sample both edges as previously. Although in the simulation studies below we did not observe false negatives (cells within $\bfT$ that falsely indicate the absence of an edge), to avoid such an eventuality being detrimental for the model search algorithm we also include completely random additions, removals or replacements within the model search. 

\section{Simulation studies}

The model used for each one of the simulations is presented in Figure 1. Simulation 1 is based on two distinct sets of variables, where variables that belong to different sets are independent.  Simulations 2 and 3 describe more complex structures compared to simulation 1, since interaction terms share variables. Simulation 4 is focused on the presence of conditional independence. We provide additional information on the design matrices and parameter coefficients of the adopted log-linear models in the Supplemental material, Section S3. 

\begin{center}
\begin{table*}[!b]
{\bf Table 1:} {Specifications for each one of the 4 simulations.}
\par
\begin{tabular}{ccccc}
\hline
Number & Number & Number of levels & Number of cells & Size of  model \\
of subjects  & of variable & of variables & in contingency table & space  \\
 10000 & 10  & 2 & 1024 & $3.5184 \times 10^{13}$ \\ 
\hline
\end{tabular}
\end{table*}
\end{center}

\begin{figure*}[!t]
\includegraphics[height=120mm, width=180mm]{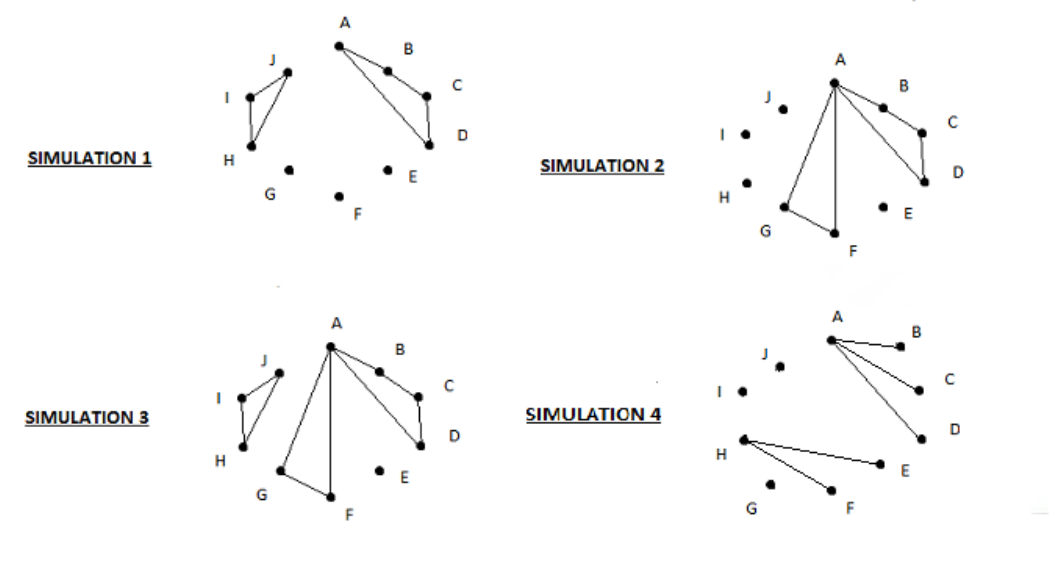}
\caption{The graphical models used for the four simulations}
\label{fig:1}       
\end{figure*}

\subsection{MCMC specifications, prior distributions and model search strategies}

The specifications for the data generating mechanism for the 4 simulations are shown in Table 1. Information on the size of the chains, as well as run times, is provided in Table 2. For the bi-clustering analyses we specified a maximum number of 8 views and 8 clusters within each view. The following prior specifications were adopted. For the clustering Dirichlet process model we adopted the non-sparse prior for $\rho_p$. 
Conjugate Dirichlet priors with $\lambda_{1}=...=\lambda_{M_{p}}=0.5$ were adopted for the $\phi_{p}^{c}$ parameters. Chains were initialized by allocating variables to 2 views using the distance based `kmodes' approach and the `klaR' R package (Huang, Z., 1997). Within each view, subjects were initially allocated to the same cluster. We found that starting with an initial view and cluster allocation where variables and subjects are allocated in such a small number of components favours faster convergence to the true undelying allocations. Initial values for all other model parameters were random. 

For the log-linear model comparison analyses, unit information priors (Ntzoufras et al., 2003) were adopted for the model parameters. All possible graphs are equally likely a priori. The log-linear models were fitted and compared using the reversible jump MCMC algorithm described in Papathomas et al. (2011). The obtained chains provide valuable information for the mixing performance of the  algorithms, although time constraints led to acquiring a number of samples that is rather small for accurately estimating  posterior probabilities of less prominent models, in model spaces as large as the ones we consider.

Following standard practice when building a reversible jump MCMC chain, in $60\%$ of the iterations a new set of values for the parameters of the currently accepted model is proposed. A jump to a different graphical model is attempted in $40\%$ of the iterations, where it is equally likely to attempt the addition, removal or replacement of one edge with another.
We compare three model search strategies for selecting the edges to add, remove or replace:
\begin{itemize}
\item[(a)] Uniformly random selection. An unrefined search strategy where all candidate edges are equally likely to be selected. 
\item[(b)] A balanced combination of (a) and the single-view based approach described in Papathomas and Richardson (2016), where the two model search approaches are each employed in $20\%$ of the iterations. We refer to this approach as `Single-view 2016'. 
\item[(c)] A balanced combination of (a) and the bi-clustering based approach described in Sections 4.2 and 4.3, where the two model search approaches are each employed in $20\%$ of the iterations. 
\end{itemize}

It is prudent to include random search steps in (b) and (c), not depended on the $\bfT$ matrix, as a safeguard, in case variable selection within clustering does not detect an existing edge in a high probability graphical model. In this scenario, the search moves that do not depend on $\bfT$ will allow for the detection of the model subspace that is not supported by the clustering. Although in all simulations and the real data analysis in Section 6.1 we did not observe edges of prominent graphical models not reflected in $\bfT$, this may happen when the signal is very weak in term of the prominent model (see Section 6.2) or low probability models in general. The majority of the specifications described above are also adopted in the real data analyses presented in Section 5, with differences indicated clearly therein.

\begin{small}
\begin{center}
\begin{table*}[!b]
\begin{small}
{\bf Table 2:} {MCMC specifications for the clustering analyses, and also for the log-linear model comparison Reversible jump chains. Analyses for the simulated data were performed on a server with Intel(R) Core(TM) i9-13900K processor (8 performance cores + 16 efficiency cores) with 128GB DDR4 RAM. Analyses for the real data application were performed on a PC equipped with an Intel(R) Core(TM) i7-1355U CPU @ 1.70GHz with 64GB RAM}
\par
\begin{tabular}{lccc}
\hline
\multicolumn{4}{c}{Clustering algorithms} \\
 & Burn-in & Iter. after burn-in (thin by 10) & Run time (hours) \\
Each simulation & 395000 & 5000  & 183 \\
CHD real data & 395000 & 5000 & 13      \\
GRACE real data & 395000 & 5000 & 13      \\
\hline
\multicolumn{4}{c}{Reversible jump chains}  \\
 & Burn-in & Iterations & Run time in hours  \\
Each simulation & 2000 & 10000  & 86 \\
CHD real data & 20000 & $10^5$  & 1   \\
GRACE real data & 20000 & $10^5$  & 1   \\
\hline
\end{tabular}
\end{small}
\end{table*}
\end{center}
\end{small}

\subsection{Simulation results: the derived $\bfT$ matrices}

The constructed $\bfT$ matrices are shown below, denoting the matrix that corresponds to simulation $j$ by 
$\bfT^{Sj}$, $j=1,2,3,4$. We chose $thres=1$ and $sumthres=0.000005$. These values were found empirically to provide a good balance between accounting for the variability in the MCMC sampling and adhering to the theoretical results in Section 4, for all simulations and real data analyses. 

We display with bold font the values of elements that correspond to an existing edge in the true model. 
In terms of surmising non-existing edges, it is clear that, overall, zero or considerably smaller weights are given to non-existing edges, compared to existing ones. Importantly, zero or very close to zero elements in the $\bfT$ matrices always correspond to a non-existing edge. They never indicate that an existing edge is absent, something that would be detrimental to a model search algorithm. 

Next to each $\bfT^{Sj}$ matrix we show the corresponding matrix derived by the single-view approach of Papathomas and Richardson (2016), denoted by $\bfT^{SV}$. We observe that the $\bfT^{SV}$ matrices tend to suggest that edges are present between distinct/independent sets of variables much more frequently than the $\bfT^{Sj}$ ones. 

Assume underlying true $0$ or $1$ elements for the $\bfT$ matrices, where $1$ corresponds to the presence of an edge according to Figure 1. The 
mean squared differences between the elements of the underlying true $\bfT$ matrices and the elements of $\bfT^{Sj}$, $j=1,2,3,4$, are $(0.045,0.021,0.077,0.097)$ respectively. The mean squared differences between the elements of the underlying true $\bfT$ matrices and the $\bfT^{SV}$ matrices for the 4 simulations are $(0.110,0.104,0.185,0.188)$, which are roughly twice as large for 3 simulations, when compared to the proposed $\bfT^{Sj}$ matrices, and almost 5 times larger for simulation 2. These results confirm the utility of bi-clustering in the detection of complex dependence structures. In the next subsection we demonstrate the utility of the bi-clustering output in formal log-linear model comparison.

\begin{tiny}
 \[
  \bfT^{s1}=\left( \begin{array}{lllllllllll}
  & A & B & C & D & E & F & G & H & I & J  \\
  A &  & {\bf .88} &  .52 & {\bf 1.00} & .00 & .00 & .00 & .00 & .00 & .00 \\
  B &  &  & {\bf .70} & 0.34 & .00 & .00 & .00 & .00 & .00 & .00 \\
  C &  &  &  & {\bf .35} & .00 & .00 & .00 & .00 & .00 & .00  \\
  D &  &	&  & 	& .00 &	.00 &	.00 &	.00	& .00 &	.00 \\
  E &  & 	&	 & 	& 	& .00 & 	.00 &	.00 & .00 & 	.00 \\
  F &  &  &  &  &  &  & .00 &	.00 & .00 &	.00 \\
  G &  &  &  &  &  &  &  & .00	& .00 &	.00 \\
  H &  &  &  &  &  &  &  &  & {\bf 0.29} & {\bf .40} \\
  I &  &  &  &  &  &  &  &  &  & {\bf .48}
  \end{array}
  \right)
	\quad
  \bfT^{SV}=\left( \begin{array}{lllllllllll}
  & A & B & C & D & E & F & G & H & I & J  \\
  A &  & {\bf .65} &  0.77 & {\bf .82} & .10 & .12 & .10 & .63 & .47 & .47 \\
  B &  &  & {\bf .46} & 0.45 & .05 & .07 & .07 & .41 & .32 & .34 \\
  C &  &  &  & {\bf .67} & .07 & .10 & .07 & .54 & .43 & .41  \\
  D &  &	&  & 	& .09 &	.14 &	.11 &	.80	& .69 &	.63 \\
  E &  & 	&	 & 	& 	& .03 & .02 &	.12 & .09 & .09 \\
  F &  &  &  &  &  &  & .03 &	.15 & .12 &	.12 \\
  G &  &  &  &  &  &  &  & .10	& .09 &	.09 \\
  H &  &  &  &  &  &  &  &  & {\bf 1.00} & {\bf .97} \\
  I &  &  &  &  &  &  &  &  &  & {\bf .90}
  \end{array}
  \right)
  \]
	\end{tiny}

\begin{tiny}
  \[
  \bfT^{s2}=\left( \begin{array}{lllllllllll}
  & A & B & C & D & E & F & G & H & I & J  \\
  A &  & {\bf 1.00} & .44 & {\bf .88} & .00 & {\bf .99} & {\bf .96} & .00 & .00 & .00 \\
  B &  &  & {\bf .95} & .38 & .00 & .24 & .31 & .00 & .00 & .00 \\
  C &  &  &  & {\bf .92} & .00 & .29 & .29 & .00 & .00 & .00  \\
  D &  &		& 	& 	& .00 &	.42 &	.28 &	.00	& .00 &	.00 \\
  E & 	& 	& 	& 	& 	& .00 &	.00 &	.00 & .00 & 	.00 \\
  F &  &  &  &  &  &  & {\bf .99}  & .00 & 	.00 &	.00 \\
  G &  &  &  &  &  &  &  & .00	& .00 &	.00 \\
  H &  &  &  &  &  &  &  &  & .00 & .00 \\
  I &  &  &  &  &  &  &  &  &  & .00
  \end{array}
  \right)
 \quad
  \bfT^{SV}=\left( \begin{array}{lllllllllll}
  & A & B & C & D & E & F & G & H & I & J  \\
  A &  & {\bf 1.00} & .45 & {\bf .94} & .09 & {\bf .82} & {\bf .79} & .11 & .11 & .16 \\
  B &  &  & {\bf .54} & .99 & .10 & .82 & .77 & .11 & .12 & .19 \\
  C &  &  &  & {\bf .54} & .05 & .33 & .31 & .05 & .06 & .11  \\
  D &  &		& 	& 	& .08 &	.76 &	.74 &	.11	& .11 &	.17 \\
  E & 	& 	& 	& 	& 	& .07 &	.06 &	.01 & .03 &	.02 \\
  F &  &  &  &  &  &  & {\bf .70}  & .10 & 	.09 &	.15 \\
  G &  &  &  &  &  &  &  & .08	& .08 &	.13 \\
  H &  &  &  &  &  &  &  &  & .02 & .03 \\
  I &  &  &  &  &  &  &  &  &  & .03
  \end{array}
  \right)
  \]
 \end{tiny}

 \begin{tiny}
 \[
  \bfT^{s3}   = \left( \begin{array}{lllllllllll}
  & A & B & C & D & E & F & G & H & I & J  \\
  A &  & {\bf .54} & .50 & {\bf .42} & .00 & {\bf .61} & {\bf .63} & .00 & .00 & .00    \\
  B &  &  & {\bf 1.00} & .39 & .00 & .24 & .36 & .00 & .00 & .00 \\
  C &  &  &  & {\bf .85} & .00 & .32 & .28 & .00 & .00 & .00  \\
  D &  &		& 	& 	& .00 &	.70 &	.38 &	.00	& .00 &	.00 \\
  E & 	& 	& 	& 	& 	& .00 & 	.00 &	.00 & .00 &	.00 \\
  F &  &  &  &  &  &  & {\bf .90} &	.00 &	.00 &	.00 \\
  G &  &  &  &  &  &  &  & .00	& .00 &	.00 \\
  H &  &  &  &  &  &  &  &  & {\bf .40} & {\bf .22} \\
  I &  &  &  &  &  &  &  &  &  & {\bf .54}
  \end{array}
  \right)
	\quad
  \bfT^{SV}   = \left( \begin{array}{lllllllllll}
  & A & B & C & D & E & F & G & H & I & J  \\
  A &  & {\bf .70} & .59 & {\bf 1.00} & .17 & {\bf .82} & {\bf .67} & .57 & .48 & .64    \\
  B &  &  & {\bf .45} & .73 & .11 & .49 & .40 & .41 & .36 & .45 \\
  C &  &  &  & {\bf .65} & .10 & .40 & .30 & .38 & .26 & .41  \\
  D &  &		& 	& 	& .18 &	.74 &	.61 &	.66	& .52 &	.67 \\
  E & 	& 	& 	& 	& 	& .15 & 	.13 &	.13 & .12 &	.14 \\
  F &  &  &  &  &  &  & {\bf .71} &	.54 &	.46 &	.59 \\
  G &  &  &  &  &  &  &  & .53	& .48 &	.58 \\
  H &  &  &  &  &  &  &  &  & {\bf .67} & {\bf .79} \\
  I &  &  &  &  &  &  &  &  &  & {\bf .69}
  \end{array}
  \right)
	\]
 \end{tiny}

  \begin{tiny}
 \[
  \bfT^{s4}   = \left( \begin{array}{lllllllllll}
  & A & B & C & D & E & F & G & H & I & J  \\
  A &  & {\bf .37} & {\bf .33} & {\bf .18} & .00 & .00 & .00 & .00 & .00 & .00 \\
  B &  &  & 1.00 & .62 & .00 & .00 & .00 & .00 & .00 & .00 \\
  C &  &  &  & .65 & .00 & .00 & .00 & .00 & .00 & .00  \\
  D &  &		& 	& 	& .00 &	.00 &	.00 &	.00	& .00 &	.00 \\
  E & 	& 	& 	& 	& 	& .52 & .00 &	{\bf.53} & .00 & .00 \\
  F &  &  &  &  &  &  & .00 &	{\bf .26} & 	.00 &	.00 \\
  G &  &  &  &  &  &  &  & .00	& .00 &	.00 \\
  H &  &  &  &  &  &  &  &  & .00 & .00 \\
  I &  &  &  &  &  &  &  &  &  & .00
  \end{array}
  \right)
	\quad
  \bfT^{SV}   = \left( \begin{array}{lllllllllll}
  & A & B & C & D & E & F & G & H & I & J  \\
  A &  & {\bf .69} & {\bf .97} & {\bf .81} & .58 & .67 & .13 & .51 & .14 & .09 \\
  B &  &  & .80 & .68 & .42 & .49 & .10 & .41 & .11 & .07 \\
  C &  &  &  & .96 & .84 & 1.00 & .17 & .82 & .17 & .13  \\
  D &  &		& 	& 	& .59 &	.66 &	.12 &	.53	& .14 &	.10 \\
  E & 	& 	& 	& 	& 	& .90 & .15 &	{\bf.79} & .14 & .11 \\
  F &  &  &  &  &  &  & .17 &	{\bf .85} & 	.14 &	.12 \\
  G &  &  &  &  &  &  &  & .16	& .03 &	.02 \\
  H &  &  &  &  &  &  &  &  & .13 & .10 \\
  I &  &  &  &  &  &  &  &  &  & .03
  \end{array}
  \right)
	\]
 \end{tiny}

\subsubsection{Simulation results: log-linear model selection}

To assess the efficiency of the model search algorithm, we run the reversible jump chain 20 times for each of the 4 simulations, and recorded the number of iterations required to reach the true model, starting from the model that only contains main effects. We report the median, quartiles, and minimum and maximum observed number of iterations to true model. We also report the observed acceptance rate. Under the MCMC specifications in Table 2, the posterior probabilities for the true underling model for Simulations 1 to 4 are estimated as $(0.52,0.54,0.34,0.47)$.

In Table 3, we present results from the different reversible jump chains and three search strategies. The simulations show the benefit of a search strategy where information from variable selection within bi-clustering is included in log-linear model search. In terms of iterations required to reach the best model, the proposed bi-clustering approach outperforms the other two search strategies, with the completely random model search performing worse than the other two. 

There are no large differences between the three search strategies in terms of acceptance rate, although there is a downwards trend from the completely random search to the one based on bi-clustering. We observe a trade-off between acceptance rate and number of iterations to the best model. The intuitive interpretation is that with more targeted
moves the overall chance of jumping decreases, but the chain moves more quickly to the high posterior probability region.

Finally, Papathomas and Richardson (2016) demonstrated the utility of variable selection within the simpler single-view clustering model for detecting variables that are independent of all others. Our approach also reveals such variables by placing them in different views over the MCMC sampling. For example, this is the case for variables $\{E,F,G,H\}$ in Simulation 1. 

\section{Real data analyses}
\subsection{Risk factors for coronary heart disease (CHD)}

Edwards and Havr\'anek (1985) presented a $2^{6}$ contingency table
in which $1841$ men were cross-classified by six risk factors for
coronary heart disease. We assume
that main effects are always present and compare the $32768$ possible graphical log-linear
models. Due to the large number of times this data set has been analysed in the past [see, for example, Dellaportas and Forster (1999)] the top two graphical models (`ADE+AC+BC+BE+F' and `AE+DE+AC+BC+BE+F', following the notation in Agresti, 2002) and associated posterior model probabilities (0.28 and 0.23 respectively for unit information parameter priors) are known. All other graphical models have posterior probabilities lower than 0.1.

For both real data analyses, MCMC specifications as well as run times, are given in Table 2. Prior distributions were the same as the ones adopted in the analysis of the simulated data, described in Section 4.2. To assess the efficiency of the model search algorithm, we run the reversible jump chain 50 times, and recorded the number of iterations required to reach the highest probability model. 
 
The two-way interactions `AC', `AE', `BC', `BE' and `DE', that are present in the highest probability model, are clearly captured by $\bfT^{CHD}$; see below. Element $t(1,4)=0.47$ completes the three-way interaction `ADE'.  Elements that correspond to edges in high probability models are at least one order of magnitude larger compared to the five elements that correspond to `F'. Factor `F' does not interact with any other variable, and this matches its presence in a different view for many iterations, something reflected in $\bfT^{CHD}$. Using $\bfT^{CHD}$ to inform the model search algorithm, results in the identification of a large part of the model space that is associated with low probability. For comparison, we also provide below matrix $\bfT^{CHD,SV}$ created by the single-view clustering approach. The two matrices are quite similar, as the first 5 variables are placed in the same view by the bi-clustering, which captures notably better the `AD' and `DE' edges. 

In Table 4, we present model search results. It is clear that adopting the search strategy that incorporates information from bi-clustering reduces the average number of iterations to the best model to a larger extent compared to single-view clustering or a completely random search.

\begin{footnotesize}
 \[
  \bfT^{\mbox{CHD}}   = \left( \begin{array}{lllllll}
    & A & B & C & D & E & F \\
 A & & .94 & {\bf .95} & {\bf .43} & {\bf .47} & .14    \\
 B & &  & {\bf 1.00} & .33 & {\bf .51} & .07  \\
 C & &  &  & .35 & .43 & .06   \\
 D & &		& 	& 	& {\bf .60} &	.12  \\
 E &	& 	& 	& 	& 	& .26
  \end{array}
  \right)
  \quad
  \bfT^{\mbox{CHD,SV}}   = \left( \begin{array}{lllllll}
    & A & B & C & D & E & F \\
 A & & .81 & {\bf .81} & {\bf .14} & {\bf .56} & .04    \\
 B & &  & {\bf 1.00} & .16 & {\bf .75} & .05  \\
 C & &  &  & .16 & .75 & .05   \\
 D & &		& 	& 	& {\bf .12} &	.01  \\
 E &	& 	& 	& 	& 	& .05
  \end{array}
  \right)
  \]
 \end{footnotesize}

\subsection{GRACE data}

The GRACE dataset (Hosmer et al.,2008) is available from R in 
\begin{verbatim}
https://mlr3proba.mlr-org.com/reference/grace.html?q=grace#ref-usage}. 
\end{verbatim}
It contains clinical measurements for 1000 subjects, as well as time-to-event and follow-up time. We include in our analysis:
\begin{itemize}
\item A, Follow-up-time (0-less or equal to 20 days, 1-greater than 20 days)
\item B, Death (0-death, 1-censored)
\item C, Revascularisation (0-not performed, 1-performed)
\item D, Revascularisation days (0-revascularisation performed less than or equal to 20 days since admission, 1-more than 20 days or no revascularisation)
\item E, Systolic blood pressure (0-less or equal to the median 140  mm Hg, greater than 140  mm Hg)
\item F, ST-segment deviation on index ECG (0-No, 1-yes)
\end{itemize}

0.5 was added to 0 cell-counts in the contingency table to avoid numerical instability. In such analyses it is advisable to also include age in the model. As age relates to almost all measured variables, its inclusion creates a connected graph, and all variables are placed in a single view. To allow demonstrate aspects of our approach, age was excluded from this analysis. 

The top four graphical models are `ABC+ACD+BE+BF', `ABC+ACD+AE+BF', \\
`ABD+ACD+BE+BF', and `ABD+ACD+AE', with associated posterior model probabilities 0.16, 0.14, 0.13 and 0.12 respectively for unit information priors, with the remaining models with probabilities at least one order of magnitude smaller.

The $\bfT$ matrices below indicate with bold the edges present in all four highest probability models. Both matrices capture these edges. Because there are four prevalent models of almost equal and quite low probability, it is not clear which matrix captures better the underlying truth. Edge `AB' is not captured as well by the bi-clustering approach, but the single-view approach gives a much larger weight to edge `BD' not present in the best model. Because of the low posterior model probabilities, the signal is not strong in terms of edges that should be prevalent, other than the ones that are present in all four top models. This demonstrates the benefit of including random search moves in addition to utilising the $\bfT$ matrix. Even in the presence of such weak signal, the model selection results in Table 4 show that utilizing information from bi-clustering reduces markedly the average number of iterations to the best model compared to single-view clustering or a completely random search. 

\begin{footnotesize}
 \[
  \bfT^{\mbox{GRACE}}   = \left( \begin{array}{lllllll}
    & A & B & C & D & E & F \\
 A & &  {\bf .15} & {\bf 0.54} & {\bf 1.00} & .00 & .02    \\
 B & &  & .14 & .15 & .04 & .07  \\
 C & &  &  & {\bf .56} & .17 & .02   \\
 D & &		& 	& 	& .00 &	.02  \\
 E &	& 	& 	& 	& 	& .05
  \end{array}
  \right)
  \quad
  \bfT^{\mbox{GRACE,SV}}   = \left( \begin{array}{lllllll}
    & A & B & C & D & E & F \\
 A & & {\bf .78} & {\bf 1.00} & {\bf 1.00} & .20 & .09    \\
 B & &  & .77 & .78 & .16 & .06  \\
 C & &  &  & {\bf .99} & .19 & .09   \\
 D & &		& 	& 	& .19 &	.09  \\
 E &	& 	& 	& 	& 	& .02
  \end{array}
  \right)
  \]
 \end{footnotesize}

\section{Discussion}

In this manuscript, the theoretical results in Section 4 establish that bi-clustering outcomes  inform on the dependence structure of categorical variables in a more holistic manner compared to single-view modelling. The analysis of simulated and real data showed that a log-linear model search algorithm is more efficient in terms of reaching high-probability areas when informed by bi-clustering outcomes, compared to utilising single-view outcomes or adopting completely random jumps. 

Note that bi-clustering results are only used to assist the model search algorithm, and not to form the log-linear modelling prior distributions. So, the data are not used twice and there is no adverse effect regarding model selection and the evaluation of uncertainty. The prior distributions adopted for bi-clustering and log-linear modelling are largely non-informative. Nevertheless, as future work, it would be of interest to investigate if the two priors match or how they should be formed to effect prior matching. Any mismatch between the prior distributions may be an additional reason that explains bi-clustering results not translating exactly to the variables' dependence structure described by the graphical log-linear model. 

We observed through empirical results that variables that do not contribute to any clustering tend to form an individual view. This means that if many variables are each one independent of all others, then a large maximum number of potential views would be required. This would result in a clustering MCMC algorithm that is very demanding computationally. This issue can be addressed easily by identifying and removing such variables following the approach of Papathomas and Richardson (2016).  

It is of interest to consider how the inherent randomness within the MCMC sampling, including possible lack of convergence, may affect the derived $\bfT$ matrices. To this end, and for Simulation 1, we run an MCMC bi-clustering chain fixing the view allocation to the real one, producing matrix $\bfT^{FixedViews}$ As shown below, results in terms of the derived matrix are similar to the ones when the view allocation is unknown and sampled. The weight placed on edges is not the same, but it is demonstrably not zero for the same elements of the two matrices.  

\begin{tiny}
 \[
  \bfT^{s1}=\left( \begin{array}{lllllllllll}
  & A & B & C & D & E & F & G & H & I & J  \\
  A &  & {\bf .88} &  .52 & {\bf 1.00} & .00 & .00 & .00 & .00 & .00 & .00 \\
  B &  &  & {\bf .70} & 0.34 & .00 & .00 & .00 & .00 & .00 & .00 \\
  C &  &  &  & {\bf .35} & .00 & .00 & .00 & .00 & .00 & .00  \\
  D &  &	&  & 	& .00 &	.00 &	.00 &	.00	& .00 &	.00 \\
  E &  & 	&	 & 	& 	& .00 & 	.00 &	.00 & .00 & 	.00 \\
  F &  &  &  &  &  &  & .00 &	.00 & .00 &	.00 \\
  G &  &  &  &  &  &  &  & .00	& .00 &	.00 \\
  H &  &  &  &  &  &  &  &  & {\bf 0.29} & {\bf .40} \\
  I &  &  &  &  &  &  &  &  &  & {\bf .48}
  \end{array}
  \right)
	\]
	\[
  \bfT^{FixedViews}=\left( \begin{array}{lllllllllll}
  & A & B & C & D & E & F & G & H & I & J  \\
  A &  & {\bf .58} &  .62 & {\bf .69} & .00 & .00 & .00 & .00 & .00 & .00 \\
  B &  &  & {\bf .45} & 0.55 & .00 & .00 & .00 & .00 & .00 & .00 \\
  C &  &  &  & {\bf .65} & .00 & .00 & .00 & .00 & .00 & .00  \\
  D &  &	&  & 	& .00 &	.00 &	.00 &	.00	& .00 &	.00 \\
  E &  & 	&	 & 	& 	& .00 & 	.00 &	.00 & .00 & 	.00 \\
  F &  &  &  &  &  &  & .00 &	.00 & .00 &	.00 \\
  G &  &  &  &  &  &  &  & .00	& .00 &	.00 \\
  H &  &  &  &  &  &  &  &  & {\bf 1.00} & {\bf .76} \\
  I &  &  &  &  &  &  &  &  &  & {\bf .68}
  \end{array}
  \right)
  \]
	\end{tiny}
	
	As seen in Table 2, implementing the proposed approached can be computationally expensive and requires a significant amount of time. This manuscript provides the underlying theory and proof of concept analyses regarding the utility of bi-clustering. Future work will include the development of faster computational resources, to allow for the implementation of the proposed methodology to large datasets. 
	
\begin{small}
\begin{center}
\begin{table*}[!h]
\begin{tiny}
{\bf Table 3:} {Mixing performance of samplers. Median of iterations to best model is calculated after 20 runs of the reversible jump MCMC chain. Minimum, first and third quartiles and maximum values are given in parentheses in this order. Uniformly random refers to the unrefined model search strategy adopted in Papathomas et al (2011b). Single-view 2016 refers to the approach by Papathomas and Richardson (2016). Bi-clustering refers to the approach proposed in this manuscript.}
\par
\begin{tabular}{lcc}
\hline
\multicolumn{3}{c}{Simulation 1} \\
 & Acceptance rate & Iterations (median) to highest  \\
 &  as a percentage & posterior probability model   \\
(a) Uniformly random  & 5.7 & 721 (302,492,844,1592)   \\
(b) Single-view 2016 (20\%,20\%) & 5.4. &  320 (153,298,458,1167)   \\
(c) Bi-clustering (20\%,20\%) & 3.8 & 132 (36,98,220,547)   \\
\hline
\multicolumn{3}{c}{Simulation 2} \\
 & Acceptance rate & Iterations (median) to highest \\
 &  as a percentage & posterior probability model  \\
(a) Uniformly random  & 4.9 & 735 (256,546,936,1187)   \\
(b) Single-view 2016 (20\%,20\%) & 4.72 & 353 (198,284,477,734)  \\
(c) Bi-clustering (20\%,20\%) & 3.9 & 214 (53,138,295,714)   \\
\hline
\multicolumn{3}{c}{Simulation 3} \\
 & Acceptance rate & Iterations (median) to highest \\
 &  as a percentage & posterior probability model  \\
(a) Uniformly random & 4.7 & 801 (165,560,1062,1401)   \\
(b) Single-view 2016 (20\%,20\%) & 4.6 & 575 (239,366,671,953)  \\
(c) Bi-clustering (20\%,20\%) & 4.4 & 408 (256,363,571,1472)   \\
\hline
\multicolumn{3}{c}{Simulation 4} \\
 & Acceptance rate & Iterations (median) to highest \\
 &  as a percentage & posterior probability model  \\
(a) Uniformly random  & 5.6 & 541 (269,479,689,1165)  \\
(b) Single-view 2016 (20\%,20\%) & 6.3 & 509 (176,417,636,904)  \\
(c) Bi-clustering (20\%,20\%) & 4.8 & 243 (99,190,357,1388)   \\
\hline
\end{tabular}
\end{tiny}
\end{table*}
\end{center}
\end{small}

\begin{small}
\begin{center}
\begin{table*}[!h]
\begin{tiny}
{\bf Table 4:} {Mixing performance of samplers. Median of iterations to best model is calculated after 50 runs of the reversible jump MCMC chain. Minimum, first and third quartiles and maximum values are given in parentheses in this order. Uniformly random refers to the unrefined model search strategy adopted in Papathomas et al (2011b). Single-view 2016 refers to the approach by Papathomas and Richardson (2016). Bi-clustering refers to the approach proposed in this manuscript.}
\par
\begin{tabular}{lcc}
\hline
\multicolumn{3}{c}{Edwards and Havranek data (CHD)} \\
 & Acceptance rate & Iterations (median) to highest \\
 &  as a percentage & posterior probability model \\
 & &   `ADE+AC+BC+BE+F' \\
(a) Uniformly random & 5.2 & 277 (52,197,478,1364)   \\
(b) Single-view 2016 (20\%,20\%)  & 4.4 & 271 (40,185,432,1793)   \\
(c) Bi-clustering (20\%,20\%)  & 5.6 & 270 (43,145,361,2060)  \\
\hline
\multicolumn{3}{c}{GRACE data} \\
 & Acceptance rate & Iterations (median) to highest \\
 &  as a percentage & posterior probability model  \\
 & &  `ABC+ACD+BE+BF' \\
(a) Uniformly random & 5.01 & 5943 (224,1265,30884, more than 50000 on 9 occasions) \\
(b) Single-view 2016 (20\%,20\%)  & 4.9 & 2410 (136,652,21845, more than 50000 on 3 occasions)  \\
(c) Bi-clustering (20\%,20\%)  & 4.9 & 1347 (94,430,4093, more than 50000 on 1 occasion)  \\
\hline
\end{tabular}
\end{tiny}
\end{table*}
\end{center}
\end{small}

\vspace{0.2cm}
\noindent {\bf Acknowledgments}
\vspace{0.1cm}

\noindent
 We would like to thank Professor Sylvia Richardson and Dr Paul Kirk for valuable discussions during the preparation of this manuscript. 


\parindent 0mm \parskip 0cm
\vspace{0.3cm}
\noindent {\bf References}
\vspace{0.0cm}

\begin{list}{}
{\setlength{\itemsep}{0cm}
\setlength{\parsep}{0cm}
\setlength{\leftmargin}{0.5cm}
\setlength{\labelwidth}{0.5cm}
\setlength{\itemindent}{-0.5cm}
}

\item Agresti, A., 2002. Categorical data analysis. 2nd edition. John wiley \& Sons. New Jersey. 


\item Bhattacharya, A., Dunson, D.B., 2012. Simplex factor models for multivariate unordered categorical data. J. Am. Stat. Assoc. 107, 362-77.






\item D'Angelo, S., Alf\'o, M., Fop, M. 2023. Model-based clustering for multidimensional social networks, J. Roy. Stat. Soc. A, 186, 481-507.

\item Dellaportas, P., Forster, J.J., 1999. Markov chain Monte Carlo model determination for hierarchical and graphical log-linear models. Biometrika, 86, 615-633.



\item Dombowsky, A., Dunson, D.B., 2026. Product Centred Dirichlet Processes for Bayesian Multiview Clustering. J. Roy. Stat. Soc. B.  https://pmc.ncbi.nlm.nih.gov/articles/PMC12392789/


\item Dunson, D.B., Xing C., 2009. Nonparametric Bayes modelling of multivariate categorical data. J. Am. Stat. Assoc. 104, 1042-1051.

\item Edwards, D., Havr\'anek, T., 1985. A fast procedure for model search in multi-dimensional contingency tables. Biometrika, 72, 339-351.



\item Franzolini, B., De Iorio, M., Eriksson, J., 2026. 
Conditional Partial Exchangeability: A Probabilistic Framework for Multi-View Clustering,
J. Am. Stat. Assoc. DOI: 10.1080/01621459.2025.2609381

\item Fr{\"u}hwirth-Schnatter, S., Malsiner-Walli, G. and B. Gr{\"u}n (2021). Generalized mixtures of
finite mixtures and telescoping sampling. Bayesian Analysis, 16, 1279-1307.


\item Green, P.J., 1995. Reversible jump MCMC computation and Bayesian model determination. Biometrika, 82, 711-732.


\item Gr\"un, B., Malsiner-Walli, G. 2022. Bayesian finite mixture
models. In N. Balakrishnan, Theodore Colton, Brian Everitt, Walter Piegorsch, Fabrizio
Ruggeri, and Jef L. Teugels, editors, Wiley StatsRef: Statistics Reference Online doi:10.
1002/9781118445112.stat08373.


\item Hosmer, D.W., Lemeshow, S., May, S. 2008. Applied Survival Analysis: Regression Modeling of Time to Event Data: Second Edition, John Wiley and Sons Inc., New York, NY

\item Huang, Z. (1997) A Fast Clustering Algorithm to Cluster Very Large Categorical Data Sets in Data Mining. in KDD: Techniques and Applications (H. Lu, H. Motoda and H. Luu, Eds.), pp. 21-34, World Scientific, Singapore.



\item Ishwaran, H.,James, L., 2001. Gibbs sampling methods for stick-breaking priors. J. Am. Stat. Assoc. 96, 161-73.

\item Jensen AB, Moseley PL, Oprea TI, et al. 2014. Temporal disease trajectories condensed from population-wide registry data covering 6.2 million patients. Nature Communications, 5(1): 4022. DOI: 10.1038/ncomms5022.

\item Johndrow, J.E., Bhattacharya, A., Dunson, D.B., 2014. Tensor decompositions and sparse log-linear models. arXiv:1404.0396v1. 

\item Keenan, K., Papathomas M., et al. 2024. Intersecting social and environmental determinants of multidrug-resistant urinary tract infections in East Africa beyond antibiotic use. Nature Communications. 15, 11 p.

\item Kirk, P., Pagani, F. Richardson, S. 2023. Bayesian outcome-guided multi-view mixture models with applications in molecular precision medicine. 	arXiv:2303.00318


\item Lauritzen, S.L., 2011. Elements of graphical models. Lectures from the XXXVIth International Probability Summer School in St-Flour, France. http://www.stats.ox.ac.uk/~steffen

\item Liverani, S., Hastie, D. I., Azizi, L., Papathomas, M., Richardson, S., 2015. PReMiuM: An R package for Profile Regression Mixture Models using Dirichlet Processes. Journal of Statistical Software, 64(7), 1-30.   



\item Mantziou, A., Lunag\'omez, S., Mitra, R. 2024. Bayesian model-based clustering for populations of network data. Ann. Appl. Stat., 18, 266-302.

\item Marbac, M., Biernacki, C., Vandewalle, V., 2014. Model-based clustering for conditionally correlated categorical data. arXiv:1401.5684v2


\item Ntzoufras, I., Dellaportas, P., Forster, J.J., 2003. Bayesian variable and link determination for generalized linear models. J. Stat. Plan. Infer. 111, 165-180.



\item Papathomas, M., Dellaportas, P., Vasdekis, V.G.S., 2011b. A novel reversible jump algorithm for generalized linear models. Biometrika, 98, 231-236.

\item Papathomas, M., Molitor, J., Hoggart, C., Hastie, D., Richardson, S., 2012. Exploring data from genetic association studies using Bayesian variable selection and the Dirichlet process: application to searching for gene-gene patterns. Genet. Epidemiol. 36, 663-674.

\item Papathomas, M., Richardson, S., 2016. Exploring dependence between categorical variables: Benefits and limitations of using variable selection within Bayesian clustering in relation to log-linear modelling with
interaction terms. Journal of Statistical Planning and Inference. 173, 47-63.







\item West, M., 1992. Hyperparameter estimation in Dirichlet process mixture models. Institute of Statistics and Decision Sciences.


\item Zhou, J., Bhattacharya, A., Herring, A.H., Dunson, D.B., 2015. Bayesian factorizations of big sparse tensors. J. Amer. Statist. Assoc. 110, 1562-1576

\end{list}

\vspace{0.5cm}

\noindent {\bf Appendix.}
\vspace{0.1cm}
{\it Proof of Theorem 1:} To simplify the notation, we suppress the $x$ and $x^{'}$ from $P(x_{.p}=x, x_{.q}=x^{'})$, and write $P(x_{.p}, x_{.q})$. We also write $\phi_p^c$ instead of $\phi_{p}^{c}(x)$, and $\pi_p$ instead of $\pi_p(x)$. Without any loss of generality, assume that $v=1$ and $v^{'}=2$. Then, 
\[
P(x_{.p_{1}}, \dots ,x_{.p_{2}},x_{.q_{1}},\dots,x_{.q_{2}}) = 
\sum_{c^{1}=1}^{C^{1}} \sum_{c^{2}=1}^{C^{2}} 
\psi^{1}_{c^{1}} \times \psi^{2}_{c^{2}} 
\]
\[
\times 
\{(\phi_{p_{1}}^{c^{1}})^{\gamma_{p_{1}}^{c^{1}}} (\pi_{p_{1}})^{1-\gamma_{p_{1}}^{c^{1}}} \} \times \dots \times
\{(\phi_{p_{2}}^{c^{1}})^{\gamma_{p_{2}}^{c^{1}}} (\pi_{p_{2}})^{1-\gamma_{p_{2}}^{c^{1}}} \} 
\times 
\{(\phi_{q_{1}}^{c^{2}})^{\gamma_{q_{1}}^{c^{2}}} (\pi_{q_{1}})^{1-\gamma_{q_{1}}^{c^{2}}} \times \dots \times
\{(\phi_{q_{2}}^{c^{2}})^{\gamma_{q_{2}}^{c^{2}}} (\pi_{q_{2}})^{1-\gamma_{q_{2}}^{c^{2}}} \}
\} 
\]

Also, 
\[
P(x_{.p_{1}}, \dots ,x_{.p_{2}}) \times P(x_{.q_{1}},\dots,x_{.q_{2}}) 
\]
\[
= \sum_{c^{1}=1}^{C^{1}} \psi^{1}_{c^{1}}
\{(\phi_{p_{1}}^{c^{1}})^{\gamma_{p_{1}}^{c^{1}}} (\pi_{p_{1}})^{1-\gamma_{p_{1}}^{c^{1}}} \} \times \dots \times
\{(\phi_{p_{2}}^{c^{1}})^{\gamma_{p_{2}}^{c^{1}}} (\pi_{p_{2}})^{1-\gamma_{p_{2}}^{c^{1}}} \} 
\]
\[
\times 
\sum_{c^{2}=1}^{C^{2}} \psi^{2}_{c^{2}}
\{(\phi_{q_{1}}^{c^{2}})^{\gamma_{q_{1}}^{c^{2}}} (\pi_{q_{1}})^{1-\gamma_{q_{1}}^{c^{2}}} \times \dots \times
\{(\phi_{q_{2}}^{c^{2}})^{\gamma_{q_{2}}^{c^{2}}} (\pi_{q_{2}})^{1-\gamma_{q_{2}}^{c^{2}}} \}
\]

As,
\[
\sum_{k=1}^{n} \sum_{j=1}^{m} a_{k} b_{j} = \left( \sum_{k=1}^{n} a_{k} \right) 
\left( \sum_{j=1}^{m} b_{j} \right),
\]
we have,
\[
P(x_{.p_{1}}, \dots ,x_{.p_{2}},x_{.q_{1}},\dots,x_{.q_{2}}) 
\]
\[
= \sum_{c^{1}=1}^{C^{1}} \psi^{1}_{c^{1}}
\{(\phi_{p_{1}}^{c^{1}})^{\gamma_{p_{1}}^{c^{1}}} (\pi_{p_{1}})^{1-\gamma_{p_{1}}^{c^{1}}} \} \times \dots \times
\{(\phi_{p_{2}}^{c^{1}})^{\gamma_{p_{2}}^{c^{1}}} (\pi_{p_{2}})^{1-\gamma_{p_{2}}^{c^{1}}} \} 
\]
\[
\times 
\sum_{c^{2}=1}^{C^{2}} \psi^{2}_{c^{2}}
\{(\phi_{q_{1}}^{c^{2}})^{\gamma_{q_{1}}^{c^{2}}} (\pi_{q_{1}})^{1-\gamma_{q_{1}}^{c^{2}}} \times \dots \times
\{(\phi_{q_{2}}^{c^{2}})^{\gamma_{q_{2}}^{c^{2}}} (\pi_{q_{2}})^{1-\gamma_{q_{2}}^{c^{2}}} \}
\]
\[
= P(x_{.p_{1}}, \dots ,x_{.p_{2}}) \times P(x_{.q_{1}},\dots,x_{.q_{2}}).
\]
This proves Theorem 1. 

\vspace{0.1cm}
{\it Proof of Theorem 2:} \underline{Different views $\Rightarrow$ Independence}

Consider the representative $x_{.p_{1}}$ from the set of variables $\{x_{.p_{1}},\dots,x_{.p_{2}} \}$. 
so that, $v_{p_{1}} \neq v_{q}$, for $q \in \{q_{1}, \dots , q_{2} \}$. Without loss of generality, assume that $\{ v_{1}, \dots , v_{d} \}$ represent the distinct set of values from $\{ v_{q_{1}}, \dots , v_{q_{2}} \}$. Then, 
\[
P(x_{.p_{1}}, x_{.q_{1}}, \dots ,x_{.q_{2}} | \phi, \psi, \xi) = 
\sum_{c^{v_{p_{1}}}=1}^{C^{v_{p_{1}}}} \sum_{c^{v_{1}}=1}^{C^{v_{1}}} \dots \sum_{c^{v_{d}}=1}^{C^{v_{d}}} 
\]
\[
\psi^{v_{p_{1}}}_{c^{v_{p_{1}}}}
\{(\phi_{p_{1}}^{c^{v_{p_{1}}}})^{\gamma_{p_{1}}^{c^{v_{p_{1}}}}} (\pi_{p_{1}})^{1-\gamma_{p_{1}}^{c^{v_{p_{1}}}}} \} 
\times 
\psi^{v_{1}}_{c^{v_{1}}} \prod_{q:v_{q}=v_{1}} 
\{(\phi_{q}^{c^{v_{1}}})^{\gamma_{q}^{c^{v_{1}}}} (\pi_{q})^{1-\gamma_{q}^{c^{v_{1}}}} \}
\]
\[
\times ... \times  
\psi^{v_{d}}_{c^{v_{d}}} \prod_{q:v_{q}=v_{d}} 
\{(\phi_{q}^{c^{v_{d}}})^{\gamma_{q}^{c^{v_{d}}}} (\pi_{q})^{1-\gamma_{q}^{c^{v_{d}}}} \}
\]

As,
\[
\sum_{k=1}^{n} \sum_{j=1}^{m} a_{k} b_{j} = \left( \sum_{k=1}^{n} a_{k} \right) 
\left( \sum_{j=1}^{m} b_{j} \right),
\]
we have,
\[
P(x_{.p_{1}}, x_{.q_{1}}, \dots ,x_{.q_{2}} | \phi, \psi, \xi) = 
\sum_{c^{v_{p_{1}}}=1}^{C^{v_{p_{1}}}} \psi^{v_{p_{1}}}_{c^{v_{p_{1}}}}
\{(\phi_{p_{1}}^{c^{v_{p_{1}}}})^{\gamma_{p_{1}}^{c^{v_{p_{1}}}}} (\pi_{p_{1}})^{1-\gamma_{p_{1}}^{c^{v_{p_{1}}}}} \} 
\]
\[
\sum_{c^{v_{1}}=1}^{C^{v_{1}}} \dots \sum_{c^{v_{d}}=1}^{C^{v_{d}}}
\psi^{v_{1}}_{c^{v_{1}}} \prod_{q:v_{q}=v_{1}} 
\{(\phi_{q}^{c^{v_{1}}})^{\gamma_{q}^{c^{v_{1}}}} (\pi_{q})^{1-\gamma_{q}^{c^{v_{1}}}} \} \dots
\psi^{v_{d}}_{c^{v_{d}}} \prod_{q:v_{q}=v_{d}} 
\{(\phi_{q}^{c^{v_{d}}})^{\gamma_{q}^{c^{v_{d}}}} (\pi_{q})^{1-\gamma_{q}^{c^{v_{d}}}} \}
\]    
\[
= P(x_{.p_{1}}| \phi, \psi, \xi) \times P(x_{.q_{1}}, \dots ,x_{.q_{2}} | \phi, \psi, \xi)
\]  
Therefore, $x_{p_1}$, a representative for any element from $\{x_{.p_{1}},\dots,x_{.p_{2}} \}$, is independent of $\{ x_{.q_{1}}, \dots ,x_{.q_{2}} \}$. 

\underline{Independence $\Rightarrow$ Different views}

Consider the representative $x_{.p_{1}}$ from the set of variables $\{x_{.p_{1}},\dots,x_{.p_{2}} \}$. 
Assume that $x_{.p_{1}}$ is independent of $\{ x_{.q_{1}}, \dots , x_{.q_{2}} \}$. 
Assume that $\{ v_{1}, \dots , v_{d} \}$ represent the distinct 
set of values from $\{ v_{q_{1}}, \dots , v_{q_{2}} \}$. Then, 

\[
P(x_{.p_{1}}, x_{.q_{1}}, \dots ,x_{.q_{2}} | \phi, \psi, \xi) = 
P(x_{.p_{1}}| \phi, \psi, \xi) \times P(x_{.q_{1}}, \dots ,x_{.q_{2}} | \phi, \psi, \xi)
\]
\[
= \sum_{c^{v_{p_{1}}}=1}^{C^{v_{p_{1}}}} \psi^{v_{p_{1}}}_{c^{v_{p_{1}}}}
\{(\phi_{p_{1}}^{c^{v_{p_{1}}}})^{\gamma_{p_{1}}^{c^{v_{p_{1}}}}} (\pi_{p_{1}})^{1-\gamma_{p_{1}}^{c^{v_{p_{1}}}}} \} 
\]
\begin{eqnarray}
\label{Th2}
\times 
\sum_{c^{v_{1}}=1}^{C^{v_{1}}} \dots \sum_{c^{v_{d}}=1}^{C^{v_{d}}}
\psi^{v_{1}}_{c^{v_{1}}} \prod_{q:v_{q}=v_{1}} 
\{(\phi_{q}^{c^{v_{1}}})^{\gamma_{q}^{c^{v_{1}}}} (\pi_{q})^{1-\gamma_{q}^{c^{v_{1}}}} \} \dots
\psi^{v_{d}}_{c^{v_{d}}} \prod_{q:v_{q}=v_{d}} 
\{(\phi_{q}^{c^{v_{d}}})^{\gamma_{q}^{c^{v_{q}}}} (\pi_{q})^{1-\gamma_{q}^{c^{v_{d}}}} \}.
\end{eqnarray}

Now, assume that one or more variables in $\{q_{1}, \dots, q_{2} \}$ are allocated to the same view as $p_{1}$. Without loss of generality, assume that $v_{p_{1}}=v_{1}$. Then, 
\[
P(x_{.p_{1}}, x_{q_{1}}, \dots ,x_{.q_{2}} | \phi, \psi, \xi) = 
P(x_{.p_{1}}| \phi, \psi, \xi) \times P(x_{.q_{1}}, \dots ,x_{.q_{2}} | \phi, \psi, \xi)
\]
\[
=   \sum_{c^{1}=1}^{C^{1}} \psi^{1}_{c^{1}}
\{(\phi_{p_{1}}^{c^{1}})^{\gamma_{p_{1}}^{c^{1}}} (\pi_{p_{1}})^{1-\gamma_{p_{1}}^{c^{1}}} \} 
\sum_{c^{1}=1}^{C^{1}} \sum_{c^{v_{2}}=1}^{C^{v_{2}}} \dots \sum_{c^{v_{d}}=1}^{C^{v_{d}}}
\psi^{1}_{c^{1}} \prod_{q:v_{q}=1} 
\{(\phi_{q}^{c^{1}})^{\gamma_{q}^{c^{1}}} (\pi_{q})^{1-\gamma_{q}^{c^{1}}} \} 
\]

\[
\times 
\psi^{v_{2}}_{c^{v_{2}}} \prod_{q:v_{q}=v_{2}} 
\{(\phi_{q}^{c^{v_{2}}})^{\gamma_{q}^{c^{v_{2}}}} (\pi_{q})^{1-\gamma_{q}^{c^{v_{2}}}} \} \times \dots \times 
\psi^{v_{d}}_{c^{v_{d}}} \prod_{q:v_{q}=v_{d}} 
\{(\phi_{q}^{c^{v_{d}}})^{\gamma_{q}^{c^{v_{d}}}} (\pi_{q})^{1-\gamma_{q}^{c^{v_{d}}}} \}.
\]

For this expression to be equal to (\ref{Th2}) in general, we need, 
\[
\sum_{c^{1}=1}^{C^{1}} \psi^{1}_{c^{1}} 
\{(\phi_{p_{1}}^{c^{1}})^{\gamma_{p_{1}}^{c^{1}}} (\pi_{p_{1}})^{1-\gamma_{p_{1}}^{c^{1}}} \} 
\prod_{q:v_{q}=1} 
\{(\phi_{q}^{c^{1}})^{\gamma_{q}^{c^{1}}} (\pi_{q})^{1-\gamma_{q}^{c^{1}}} \}
\]
\[
=   \sum_{c^{1}=1}^{C^{1}} \psi^{1}_{c^{1}}
\{(\phi_{p_{1}}^{c^{1}})^{\gamma_{p_{1}}^{c^{1}}} (\pi_{p_{1}})^{1-\gamma_{p_{1}}^{c^{1}}} \}
\sum_{c^{1}=1}^{C^{1}} \psi^{1}_{c^{1}} \prod_{q:v_{q}=1} 
\{(\phi_{q}^{c^{1}})^{\gamma_{q}^{c^{1}}} (\pi_{q})^{1-\gamma_{q}^{c^{1}}} \} 
\]

This is not possible as, in general, $\psi^{1}_{c^{1}} \neq (\psi^{1}_{c^{1}})^2$. As we assumed at the start that $x_{.p_{1}}$ is independent of $\{ x_{.q_{1}}, \dots , x_{.q_{2}} \}$, it is not possible that one or more variables in $\{q_{1}, \dots, q_{2} \}$ are allocated to the same view as $p_{1}$.

{\it Proof of Proposition 1:} As this proof concerns variables within the same view, to simplify the notation, we omit the view indicators where relevant. Assume that only clusters of type $\Gamma_1$ and $\Gamma_2$ exist for variables $x_{.p}$ and $x_{.q}$ that are allocated in the same view.

\begin{eqnarray}
P(x_{.p}, x_{.q}) &=& \sum_{c=1}^{C} \psi_c
\{(\phi_p^c)^{\gamma_p^c} (\pi_p)^{1-\gamma_p^c} \}
\{ (\phi_q^c)^{\gamma_q^c} (\pi_q)^{1-\gamma_q^c} \}  \nonumber \\
&=& \pi_p \sum_{c \in \Gamma_1} \psi_c \phi_q^c + \pi_q \sum_{c \in \Gamma_2} \psi_c \phi_p^c. \nonumber
\end{eqnarray}

Also,
\begin{eqnarray}
P(x_{.p}) P(x_{.q}) &=& \left( \sum_{c \in \Gamma_1} \psi_c \pi_p + \sum_{c \in \Gamma_2} \psi_c \phi_p^c  \right)
\times \left( \sum_{c \in \Gamma_1} \psi_c \phi_q^c + \sum_{c \in \Gamma_2} \psi_c \pi_q \right) \nonumber \\
&=& \left(\pi_p \sum_{c \in \Gamma_1} \psi_c  \right) \left( \sum_{c \in \Gamma_1} \psi_c \phi_q^c  \right) 
+ \left( \pi_q \sum_{c \in \Gamma_2} \psi_c  \right) \left( \sum_{c \in \Gamma_2} \psi_c \phi_p^c  \right) \nonumber \\
&+& \left( \pi_p \sum_{c \in \Gamma_1} \psi_c   \right) \left( \pi_q \sum_{c \in \Gamma_2} \psi_c  \right) 
+ \left( \sum_{c \in \Gamma_1} \psi_c \phi_q^c  \right)  \left( \sum_{c \in \Gamma_2} \psi_c \phi_p^c  \right). 
\nonumber 
\end{eqnarray}

Therefore,
\[
P(x_{.p}, x_{.q}) - P(x_{.p}) P(x_{.q}) =
\pi_p \sum_{c \in \Gamma_1} \psi_c \phi_q^c \left( 1-  \sum_{c \in \Gamma_1} \psi_c  \right) 
+ \pi_q \sum_{c \in \Gamma_2} \psi_c \phi_p^c \left( 1- \sum_{c \in \Gamma_2} \psi_c \right)
\]
\[
- \pi_p \pi_q \sum_{c \in \Gamma_1} \psi_c \sum_{c \in \Gamma_2} \psi_c 
- \left( \sum_{c \in \Gamma_1} \psi_c \phi_q^c \right) \left( \sum_{c \in \Gamma_2} \psi_c \phi_p^c \right)
\]
\[
= \left( \pi_p \sum_{c \in \Gamma_2} \psi_c - \sum_{c \in \Gamma_2} \psi_c \phi_p^c \right) 
\left( \sum_{c \in \Gamma_1} \psi_c \phi_q^c -  \pi_q \sum_{c \in \Gamma_1} \psi_c  \right).
\]

This is always zero since, as $\pi_{p}(x)=P(x_{.p}=x)$,
\[
\pi_p=P(x_{.p}) = \sum_{c} \psi_c (\phi_p^c)^{\gamma_p^c} (\pi_p)^{1-\gamma_p^c} =
\pi_p \sum_{c \in \Gamma_1} \psi_c +   \sum_{c \in \Gamma_2} \psi_c \phi_p^c
\]
\[
\Rightarrow \sum_{c \in \Gamma_2} \psi_c \phi_p^c = \pi_p - \pi_p \left( 1-\sum_{c \in \Gamma_2} \psi_c \right)
\]
\[
\Rightarrow \sum_{c \in \Gamma_2} \psi_c \phi_p^c = \pi_p \sum_{c \in \Gamma_2} \psi_c .
\]
Similar calculations hold for $x_{.q}$. This means that the variables are independent, but this contradicts Theorem 2. Therefore, it is impossible to
not have a cluster in the $\Gamma_3$ or  $\Gamma_4$ category. The case where only clusters of type $\Gamma_1$ or $\Gamma_2$ exist can be shown trivially following a similar reasoning to the one above. 

\vspace{0.1cm}
{\it Proof of Theorem 3:} 
As this proof concerns variables within the same view, to simplify the notation, we omit the view indicators where relevant. 
Consider that, for $x_{.p}$ and $x_{.q}$, only clusters of type $\Gamma_3$ are present within their view $v$, i.e. $\gamma_{p}^{c}=\gamma_{q}^{c}=1$, $c=1,...,C$, so that,
\[
P(x_{.p}, x_{.q}) = \sum_{c \in \Gamma_3} \psi_c \phi_p^c \phi_q^c
\]
Now, assume that $x_{.p}$ and $x_{.q}$ are conditionally independent given the other variables in the view. We will show that this is not possible, given the initial condition that only $\Gamma_3$-type clusters are present. Denote by `$rest$' the rest of the variables within $v$. To simplify the notation, and without loss of generality, assume that the rest of the variables are $(x_{.1}, ... , x_{.r})$. Let,  
\[
f_{rest}(\phi,\pi,c) = \left( (\phi_1^c)^{\gamma_1^c} \pi_{1}^{1-\gamma_1^c} \right) \times ... \times 
\left( (\phi_r^c)^{\gamma_r^c} \pi_{r}^{1-\gamma_r^c} \right).
\]

From the conditional independence assumption,
\[
P(x_{.p}, x_{.q} | rest) = P(x_{.p} | rest) P(x_{.q} | rest) \Leftrightarrow
\] 
\[
P(x_{.p}, x_{.q}, rest) = P(x_{.p}, rest) P(x_{.q}, rest) \frac{1}{P(rest)}.
\]

This equality translates to,
\[
\sum_{c_1=1}^{C} \psi_{c_1} f_{rest}(\phi,\pi,c_1) \phi_p^{c_1} \phi_q^{c_1} = 
\frac{\left( \sum_{c_2=1}^{C} \psi_{c_2} f_{rest}(\phi,\pi,c_2) \phi_p^{c_2} \right)  
\left( \sum_{c_3=1}^{C} \psi_{c_3} f_{rest}(\phi,\pi,c_3) \phi_q^{c_3} \right)}
{ \sum_{c_4=1}^{C} \psi_{c_4} f_{rest}(\phi,\pi,c_4)}
\]
\[
\Rightarrow 
\left(  \sum_{c_1=1}^{C} \psi_{c_1} f_{rest}(\phi,\pi,c_1) \phi_p^{c_1} \phi_q^{c_1} \right) \times 
\left( \sum_{c_4=1}^{C} \psi_{c_4} f_{rest}(\phi,\pi,c_4) \right) 
\]
\[\
= \left( \sum_{c_2=1}^{C} \psi_{c_2} f_{rest}(\phi,\pi,c_2) \phi_p^{c_2} \right)  \left( \sum_{c_3=1}^{C} \psi_{c_3} f_{rest}(\phi,\pi,c_3) \phi_q^{c_3} \right)
\]
\[
\Rightarrow \sum_{c_1=1}^{C} \sum_{c_4=1}^{C} \psi_{c_1} f_{rest}(\phi,\pi,c_1) \phi_p^{c_1} \phi_q^{c_1} \psi_{c_4} f_{rest}(\phi,\pi,c_4)
\]
\[
= \sum_{c_2=1}^{C^{ne}} \sum_{c_3=1}^{C} \psi_{c_2} f_{rest}(\phi,\pi,c_2) \phi_p^{c_2} \psi_{c_3} f_{rest}(\phi,\pi,c_3) \phi_q^{c_3}.
\]

This equality does not hold in general, unless $\phi_{p}^{1}=...=\phi_{p}^{C}=\left( \sum_{\phi_{p}^{c}} \right)/C$, or $\phi_{q}^{1}=...=\phi_{q}^{C}=\left( \sum_{\phi_{q}^{c}} \right)/C$. This is not possible, because we have assumed that the relevant selection switches are equal to 1, and this implies differences in the cluster-specific $x_{.p}$ probabilities. (Similarly for $x_{.q}$.)

\vspace{0.1cm}
{\it Proof of Theorem 4:} 
As this proof concerns variables within the same view $v$, to simplify the notation, we omit the view indicators where relevant. Denote by `$rest$' the rest of the variables within $v$. 
Without any loss of generality, assume that the rest of the variables in view $v$ are $(x_{.1}, ... , x_{.r})$, and define $f_{rest}$ as in the proof for Theorem 3. 

Assume that only clusters of type $\Gamma_1$ and $\Gamma_3$ exist. (The proof follows similarly for the assumption that only clusters of type $\Gamma_1$ and $\Gamma_2$ exist.) Then, conditional independence translates to,
\[
P(x_{.p}, x_{.q} | rest) = P(x_{.p} | rest) \times P(x_{.q} | rest) 
\]
\[
\Leftrightarrow  P(x_{.p}, x_{.q}, rest) \times P(rest) = P(x_{.p}, rest) \times P(x_{.q}, rest) 
\]
\[
\Leftrightarrow  \left( \sum_{c_1 \in \Gamma_1} \psi_{c_1} f_{rest}(\phi,\pi,c_1) \pi_p \phi_q^{c_1} 
+ \sum_{c_1 \in \Gamma_3} \psi_{c_1} f_{rest}(\phi,\pi,c_1) \phi_p^{c_1}   \phi_q^{c_1}  \right)
\]
\[
\times \left( \sum_{c_4 \in \Gamma_1} \psi_{c_4} f_{rest}(\phi,\pi,c_4) + \sum_{c_4 \in \Gamma_3} \psi_{c_4} f_{rest}(\phi,\pi,c_4) \right)
\]
\[
= \left( \sum_{c_2 \in \Gamma_3} \psi_{c_2} f_{rest}(\phi,\pi,c_2) \phi_p^{c_2} 
+ \sum_{c_2 \in \Gamma_1} \psi_{c_2} f_{rest}(\phi,\pi,c_2) \pi_p  \right)
\]
\[
\times \left( \sum_{c_3 \in \Gamma_1} \psi_{c_3} f_{rest}(\phi,\pi,c_3) \phi_q^{c_3}  
+ \sum_{c_3 \in \Gamma_3} \psi_{c_3} f_{rest}(\phi,\pi,c_3) \phi_q^{c_3} \right)
\]

\[
\Leftrightarrow 
\sum_{c_1 \in \Gamma_1} \sum_{c_4 \in \Gamma_1} \psi_{c_1} \psi_{c_4} f_{rest}(\phi,\pi,c_1) f_{rest}(\phi,\pi,c_4) \pi_p \phi_q^{c_1}
+ \sum_{c_1 \in \Gamma_1} \sum_{c_4 \in \Gamma_3} \psi_{c_1} \psi_{c_4} f_{rest}(\phi,\pi,c_1) f_{rest}(\phi,\pi,c_4) \pi_p \phi_q^{c_1}  
\]
\[
+ \sum_{c_1 \in \Gamma_3} \sum_{c_4 \in \Gamma_1} \psi_{c_1} \psi_{c_4} f_{rest}(\phi,\pi,c_1) f_{rest}(\phi,\pi,c_4) \phi_p^{c_1} \phi_q^{c_1}
+ \sum_{c_1 \in \Gamma_3} \sum_{c_4 \in \Gamma_3} \psi_{c_1} \psi_{c_4} f_{rest}(\phi,\pi,c_1) f_{rest}(\phi,\pi,c_4) \phi_p^{c_1} \phi_q^{c_1}  
\]
\[
= \sum_{c_2 \in \Gamma_1} \sum_{c_3 \in \Gamma_1} \psi_{c_2} \psi_{c_3} f_{rest}(\phi,\pi,c_2) f_{rest}(\phi,\pi,c_3) \pi_p \phi_q^{c_3}
+  \sum_{c_2 \in \Gamma_1} \sum_{c_3 \in \Gamma_3} \psi_{c_2} \psi_{c_3} f_{rest}(\phi,\pi,c_2) f_{rest}(\phi,\pi,c_3) \pi_p \phi_q^{c_3}
\]
\[
\sum_{c_2 \in \Gamma_3} \sum_{c_3 \in \Gamma_1} \psi_{c_2} \psi_{c_3} f_{rest}(\phi,\pi,c_2) f_{rest}(\phi,\pi,c_3) \phi_p^{c_2} \phi_q^{c_3}
+ \sum_{c_2 \in \Gamma_3} \sum_{c_3 \in \Gamma_3} \psi_{c_2} \psi_{c_3} f_{rest}(\phi,\pi,c_2) f_{rest}(\phi,\pi,c_3) \phi_p^{c_2} \phi_q^{c_3}
\]
The expressions before and after the equality are identical, as only the dummy variables $c_1,c_2,c_3,c_4$, used as summation indexes, change. So conditional independence holds. This completes the proof. 

\vspace{0.1cm}
{\it Proof of Theorem 5:} 
As this proof concerns variables within the same view $v$, to simplify the notation, we omit the view indicators where relevant. Denote by `$rest$' the rest of the variables within $v$. Without any loss of generality, assume that the rest of the variables in view $v$ are $(x_{.1}, ... , x_{.r})$, and define $f_{rest}$ as in the proof for Theorem 3. 

Assume that $x_{.p}$ and $x_{.q}$ in view $v$ are conditionally independent. Allowing for the three types of cluster, conditional independence translates to,
\[
P(x_{.p}, x_{.q} | rest) = P(x_{.p} | rest) \times P(x_{.q} | rest) 
\]
\[
\Leftrightarrow  P(x_{.p}, x_{.q}, rest) \times P(rest) = P(x_{.p}, rest) \times P(x_{.q}, rest) 
\]
\[
\Leftrightarrow  \left(  \sum_{c_1 \in \Gamma_1} \psi_{c_1} f_{rest}(\phi,\pi,c_1) \pi_p \phi_q^{c_1} + 
\sum_{c_1 \in \Gamma_2} \psi_{c_1} f_{rest}(\phi,\pi,c_1) \phi_p^{c_1} \pi_q +
 \sum_{c_1 \in \Gamma_3} \psi_{c_1} f_{rest}(\phi,\pi,c_1) \phi_p^{c_1} \phi_q^{c_1} \right)  
\]
\[
\times \sum_{c_4 \in \Gamma_1 U \Gamma_2 U \Gamma_3} \psi_{c_4} f_{rest}(\phi,\pi,c_4) 
\]
\[
= \left(  \sum_{c_2 \in \Gamma_2 U \Gamma_3} \psi_{c_2} f_{rest}(\phi,\pi,c_2) \phi_p^{c_2} 
+ \pi_p  \sum_{c_2 \in \Gamma_1} \psi_{c_2} f_{rest}(\phi,\pi,c_2) \right)
\]
\[
\times  \left(  \sum_{c_3 \in \Gamma_1 U \Gamma_3} \psi_{c_3} f_{rest}(\phi,\pi,c_3) \phi_q^{c_3} 
+ \pi_p  \sum_{c_3 \in \Gamma_2} \psi_{c_3} f_{rest}(\phi,\pi,c_3) \right)
\]

\[
\Leftrightarrow   
\sum_{c_1 \in \Gamma_1} \sum_{c_4 \in \Gamma_1} \psi_{c_1} \psi_{c_4} f_{rest}(\phi,\pi,c_1) f_{rest}(\phi,\pi,c_4) \pi_p \phi_q^{c_1} 
+ \sum_{c_1 \in \Gamma_1} \sum_{c_4 \in \Gamma_2} \psi_{c_1} \psi_{c_4} f_{rest}(\phi,\pi,c_1) f_{rest}(\phi,\pi,c_4) \pi_p \phi_q^{c_1} 
\]
\[
+ \sum_{c_1 \in \Gamma_1} \sum_{c_4 \in \Gamma_3} \psi_{c_1} \psi_{c_4} f_{rest}(\phi,\pi,c_1) f_{rest}(\phi,\pi,c_4) \pi_p \phi_q^{c_1} 
+ \sum_{c_1 \in \Gamma_2} \sum_{c_4 \in \Gamma_1} \psi_{c_1} \psi_{c_4} f_{rest}(\phi,\pi,c_1) f_{rest}(\phi,\pi,c_4) \phi_p^{c_1} \pi_q 
\]
\[
+ \sum_{c_1 \in \Gamma_2} \sum_{c_4 \in \Gamma_2} \psi_{c_1} \psi_{c_4} f_{rest}(\phi,\pi,c_1) f_{rest}(\phi,\pi,c_4) \phi_p^{c_1} \pi_q 
+ \sum_{c_1 \in \Gamma_2} \sum_{c_4 \in \Gamma_3} \psi_{c_1} \psi_{c_4} f_{rest}(\phi,\pi,c_1) f_{rest}(\phi,\pi,c_4) \phi_p^{c_1} \pi_q 
\]
\[
+ \sum_{c_1 \in \Gamma_3} \sum_{c_4 \in \Gamma_1} \psi_{c_1} \psi_{c_4} f_{rest}(\phi,\pi,c_1) f_{rest}(\phi,\pi,c_4) \phi_p^{c_1} \phi_q^{c_1} 
+ \sum_{c_1 \in \Gamma_3} \sum_{c_4 \in \Gamma_2} \psi_{c_1} \psi_{c_4} f_{rest}(\phi,\pi,c_1) f_{rest}(\phi,\pi,c_4) \phi_p^{c_1} \phi_q^{c_1} 
\]
\[
+ \sum_{c_1 \in \Gamma_3} \sum_{c_4 \in \Gamma_3} \psi_{c_1} \psi_{c_4} f_{rest}(\phi,\pi,c_1) f_{rest}(\phi,\pi,c_4) \phi_p^{c_1} \phi_q^{c_1}
\]

\[
= \sum_{c_2 \in \Gamma_1} \sum_{c_3 \in \Gamma_1} \psi_{c_2} \psi_{c_3} f_{rest}(\phi,\pi,c_2) f_{rest}(\phi,\pi,c_3) \pi_p \phi_q^{c_3} 
+ \sum_{c_2 \in \Gamma_1} \sum_{c_3 \in \Gamma_2} \psi_{c_2} \psi_{c_3} f_{rest}(\phi,\pi,c_2) f_{rest}(\phi,\pi,c_3) \pi_p \pi_q 
\]
\[
+ \sum_{c_2 \in \Gamma_1} \sum_{c_3 \in \Gamma_3} \psi_{c_2} \psi_{c_3} f_{rest}(\phi,\pi,c_2) f_{rest}(\phi,\pi,c_3) \pi_p \phi_q^{c_3} 
+ \sum_{c_2 \in \Gamma_2} \sum_{c_3 \in \Gamma_1} \psi_{c_2} \psi_{c_3} f_{rest}(\phi,\pi,c_2) f_{rest}(\phi,\pi,c_3) \phi_p^{c_2} \phi_q^{c_3} 
\]
\[
+ \sum_{c_2 \in \Gamma_2} \sum_{c_3 \in \Gamma_2} \psi_{c_2} \psi_{c_3} f_{rest}(\phi,\pi,c_2) f_{rest}(\phi,\pi,c_3) \phi_p^{c_2} \pi_q 
+ \sum_{c_2 \in \Gamma_2} \sum_{c_3 \in \Gamma_3} \psi_{c_2} \psi_{c_3} f_{rest}(\phi,\pi,c_2) f_{rest}(\phi,\pi,c_3) \phi_p^{c_1} \phi_q^{c_3} 
\]
\[
+ \sum_{c_2 \in \Gamma_3} \sum_{c_3 \in \Gamma_1} \psi_{c_2} \psi_{c_3} f_{rest}(\phi,\pi,c_2) f_{rest}(\phi,\pi,c_3) \phi_p^{c_2} \phi_q^{c_3} 
+ \sum_{c_2 \in \Gamma_3} \sum_{c_3 \in \Gamma_2} \psi_{c_2} \psi_{c_3} f_{rest}(\phi,\pi,c_2) f_{rest}(\phi,\pi,c_3) \phi_p^{c_2} \pi_q 
\]
\[
+ \sum_{c_2 \in \Gamma_3} \sum_{c_3 \in \Gamma_3} \psi_{c_2} \psi_{c_3} f_{rest}(\phi,\pi,c_2) f_{rest}(\phi,\pi,c_3) \phi_p^{c_2} \phi_q^{c_3}
\]

\[
\Leftrightarrow 
\sum_{c_1 \in \Gamma_1} \sum_{c_4 \in \Gamma_2} \psi_{c_1} \psi_{c_4} f_{rest}(\phi,\pi,c_1) f_{rest}(\phi,\pi,c_4) \pi_p \phi_q^{c_1} 
+ \sum_{c_1 \in \Gamma_2} \sum_{c_4 \in \Gamma_1} \psi_{c_1} \psi_{c_4} f_{rest}(\phi,\pi,c_1) f_{rest}(\phi,\pi,c_4) \phi_p^{c_1} \pi_q 
\]
\begin{eqnarray}
= \sum_{c_2 \in \Gamma_1} \sum_{c_3 \in \Gamma_2} \psi_{c_2} \psi_{c_3} f_{rest}(\phi,\pi,c_2) f_{rest}(\phi,\pi,c_3) \pi_p \pi_q 
+ \sum_{c_2 \in \Gamma_2} \sum_{c_3 \in \Gamma_1} \psi_{c_2} \psi_{c_3} f_{rest}(\phi,\pi,c_2) f_{rest}(\phi,\pi,c_3) \phi_p^{c_2} \phi_q^{c_3} \label{eq_Th5}
\end{eqnarray}
Conditional independence holds when (\ref{eq_Th5}) holds. This translates to either, 
\[
\sum_{c_1 \in \Gamma_1} \psi_{c_1} f_{rest}(\phi,\pi,c_1) \phi_q^{c_1} = \sum_{c_2 \in \Gamma_1} \psi_{c_2} f_{rest}(\phi,\pi,c_2) \pi_q
\]
or 
\[
\sum_{c_4 \in \Gamma_2} \psi_{c_4} f_{rest}(\phi,\pi,c_4) \pi_p = \sum_{c_2 \in \Gamma_2} \psi_{c_2} f_{rest}(\phi,\pi,c_2) \phi_p^{c_2}.
\]

This can be written as, 
\[
\left( \sum_{c_1 \in \Gamma_1} \psi_{c_1} f_{rest}(\phi,\pi,c_1) (\phi_q^c - \pi_q) \right) 
\times \left( \sum_{c_4 \in \Gamma_2} \psi_{c_4} f_{rest}(\phi,\pi,c_4) (\pi_p - \phi_p^c) \right)   = 0.
\]

When the equation above holds, then conditional independence holds. This proves Theorem 5, after we replace the dummy $c_4$ with $c_2$.

\end{document}